\documentclass[12pt]{article}

\usepackage{fullpage}
\usepackage{amsmath}
\usepackage{amssymb}
\usepackage{bbm}
\usepackage{color}
\usepackage{ulem}

\DeclareFontFamily{OT1}{rsfs10}{}
\DeclareFontShape{OT1}{rsfs10}{m}{n}{ <-> rsfs10 }{}
\DeclareMathAlphabet{\mathscript}{OT1}{rsfs10}{m}{n}

\def\gsim{ \lower .75ex \hbox{$\sim$} \llap{\raise .27ex \hbox{$>$}} }
\def\lsim{ \lower .75ex \hbox{$\sim$} \llap{\raise .27ex \hbox{$<$}} }
\def\be{\begin{equation}}
\def\ee{\end{equation}}
\def\bea{\begin{eqnarray}}
\def\eea{\end{eqnarray}}

\newcommand{\ns}{\normalsize}

\newcommand{\kk}{{\bf k}}

\usepackage{latexsym,amsmath,amssymb,epsfig}

\usepackage{graphicx}
\usepackage{graphicx,subfigure}
\usepackage{epstopdf}
\usepackage[body={17.5cm, 21cm},right=2cm]{geometry}
\usepackage{amssymb}
\usepackage{amsmath}
\usepackage{psfrag}
\usepackage{epsfig}
\usepackage{cancel}
 \allowdisplaybreaks[4]

\usepackage[all]{xy}

\begin{document}

\begin{titlepage}

%\vspace{-3cm}

\title{
  \hfill{\ns }  \\
  %[1em]
   {\LARGE Generating Scale-Invariant Perturbations from Rapidly-Evolving Equation of State}
\\
%[1em] 
}
\author{
   Justin Khoury$^{1}$ and Paul J. Steinhardt$^{2}$
     \\
     %[0.5em]
{\ns ${}^1$ Center for Particle Cosmology, Department of Physics \& Astronomy} \\[-0.4cm]
{\ns  University of Pennsylvania, Philadelphia, PA 19104}\\
%[0.3cm]
   {\ns ${}^2$ Department of Physics \& Princeton Center for Theoretical Science} \\[-0.4cm]
{\ns Princeton University, Princeton, NJ 08544}\\[0.3cm]}

\date{}

\maketitle

\begin{abstract}
Recently, we introduced an ekpyrotic model based on a single, canonical scalar field that generates nearly scale invariant curvature fluctuations through a purely ``adiabatic mechanism" in which the background evolution is a dynamical attractor. Despite the starkly different physical mechanism for generating fluctuations, the two-point function is identical to inflation. In this paper, we further explore this concept, focusing in particular on issues of non-gaussianity and quantum corrections. We find that the degeneracy with inflation is broken at three-point level: for the 
simplest case of an exponential potential, the three-point amplitude is strongly scale dependent, resulting in a breakdown of perturbation theory on small scales.  However, we show that the perturbative breakdown can be circumvented --- and all issues raised in Linde {\it et al.} ({\tt arXiv:0912.0944}) can be addressed --- by altering the potential such that power is suppressed on small scales. The resulting range of nearly scale invariant, gaussian modes can be as much as twelve e-folds, enough  to span the scales probed by microwave background and large scale structure observations.  On smaller scales, the spectrum is not scale invariant but is observationally acceptable.  
\end{abstract}

\end{titlepage}

\section{Introduction}

There are two known cosmological phases that transform inhomogeneous and anisotropic initial conditions into a smooth and flat universe, in agreement with observations. The first is {\it inflation}~\cite{inf}, a period of accelerated expansion
occurring shortly after the big bang, which requires a component with equation of state $w< -1/3$. Alternatively, flatness and homogeneity can be achieved during an {\it ekpyrotic} phase~\cite{oldek,seiberg,oldek2,oldek3,ekandrew,oldek4}, a period of ultra-slow contraction before the big bang, driven by a stiff fluid with $w > 1$; the ekpyrotic phase also suppresses chaotic mixmaster behavior~\cite{gratton,erick,nicolis,pretorius}.  See~\cite{ekbrief,ekreview} for reviews. In both cases, phases with nearly constant $w$ can be achieved with a  single canonical scalar field with appropriately chosen potential $V(\phi)$. Whereas inflation requires $V$ to be flat and positive, an ekpyrotic phase occurs for an exponentially steep, negative potential. A fiducial example is a negative exponential potential, $V(\phi) = -V_0e^{-c\phi/M_{\rm Pl}}$, with $c\gg 1$, corresponding to $w = c^2/2 \gg 1$. 

With nearly constant $w$ and a single scalar field, inflation also generates a nearly scale invariant and gaussian spectrum for $\zeta$, the curvature perturbation on comoving hypersurfaces~\cite{zeta1,zeta2,separateuniverse}.
The same is not true for ekpyrosis. Although in Newtonian gauge the scalar field and gravitational potential fluctuations are scale invariant, these project out of $\zeta$~\cite{oldek2}. 
Barring some higher-dimensional or stringy effects near the bounce that mixes gravitational potential and curvature fluctuations ~\cite{ekandrew}, the resulting spectrum for $\zeta$ has a strong blue tilt~\cite{nicolis,robertek}. A scale invariant spectrum can be obtained with two ekpyrotic scalar fields, through an ``entropic mechanism"~\cite{lehners,riotto,finelli} that first produces entropy fluctuations and then converts them to $\zeta$~\cite{lehners,newek1,newek2,newek3,paololeonardo,koyamaek,intuitive}, as in the New Ekpyrotic scenario~\cite{newek1,newek2,newek3}. 

Another desirable property of an inflation phase with a single scalar field is that it is a dynamical attractor. On super-horizon scales, $\zeta$ measures differences in the expansion history of distant Hubble patches~\cite{separateuniverse}. Because $\zeta \approx \delta a/a \rightarrow {\rm constant}$ at long wavelengths in single-field inflation, the perturbation can be absorbed locally by a spatial diffeomorphism~\cite{weinbergzeta}. In other words, the background solution $a(t)$ is an attractor. 

Achieving scale invariance and dynamical attractor behavior in alternative scenarios has proven to be challenging.
A contracting, dust-dominated universe yields an equation for $\zeta$ that is identical to inflation~\cite{dust,latham,robertmatter}; but $\zeta$ grows outside the horizon, indicating an unstable background.
Similarly for contracting mechanisms that rely on a time-dependent sound speed~\cite{piazza}. The contracting phase in the original ekpyrotic scenario~\cite{oldek,seiberg,oldek2} is an attractor~\cite{gratton,nicolis}, but the resulting spectrum is strongly blue rather than scale invariant~\cite{robertek,nicolis}. In the entropic mechanism relying on two ekpyrotic scalar fields~\cite{lehners,riotto,finelli}, the spectrum is unambiguously scale invariant, but the entropy direction is tachyonically unstable~\cite{koyamaek}.

Recently we have proposed a counterexample, the {\it adiabatic ekpyrotic} mechanism~\cite{PRL}, in which a single canonical scalar field drives a contracting background that is a dynamical attractor and generates a scale invariant spectrum for $\zeta$. The mechanism relies on relaxing the usual assumption that the equation of state parameter $\epsilon\equiv -\dot{H}/H^2 = 3(1+w)/2$ is nearly constant, and obtains for fairly simple forms of the potential, such as a Òlifted exponentialÓ:
\be
V(\phi) = V_0\left(1-e^{-c\phi/M_{\rm Pl}}\right)\,,
\label{potent}
\ee
with $V_0 > 0$ and $c\gg 1$. The regime of interest is the transition when $\epsilon$ rises rapidly from $\epsilon\ll 1$, where
the constant term dominates, to $\epsilon \approx c^2/2 \gg 1$, where the negative exponential term dominates. During this transition, the scale factor is nearly constant, while the equation of state parameter varies rapidly as $\epsilon\sim 1/\tau^2$, where $\tau$ is conformal time. The quantity $z \equiv a(\tau)  \sqrt{2 \epsilon(\tau)}$, which determines the evolution of $\zeta$, therefore scales as
$z  \sim (-\tau)^{-1}$ --- exactly as in inflation, where $\epsilon\approx {\rm const.}$ and $a(\tau) \approx 1/(-\tau)$! The two-point function is, therefore, identical to inflation.

In fact, a recent study has shown that the only single-field cosmologies with unit sound speed that generate a scale invariant spectrum for $\zeta$ and are dynamical attractors consist of~\cite{godfrey,baumann}: 
$i)$ inflation, with $a(\tau)\sim 1/|\tau|$ and $\epsilon\approx {\rm constant}$; $ii)$ the adiabatic ekpyrotic mechanism~\cite{PRL} mentioned above,
with $\epsilon \sim 1/\tau^2$ on a slowly contracting background; and $iii)$ the adiabatic ekpyrotic mechanism on a background that first slowly expands, then slowly contracts. 
Here we focus on the contracting version of the adiabatic ekpyrotic phase; its expanding counterpart will be studied in detail elsewhere~\cite{austin}.
See~\cite{willduality} for related work. (Another counterexample proposed recently relies on a rapidly-varying, superluminal sound speed $c_s(\tau)$~\cite{csm,csk,csfedo}.
See~\cite{piazza,csearlier} for earlier related work. Even though the background is non-inflationary, $\zeta$ is amplified because the
sound horizon is shrinking.)

The aim of this paper is to further explore the phenomenological implications of the adiabatic ekpyrotic mechanism, focusing in particular on non-gaussianities and strong coupling.
We show that the degeneracy with inflation is broken by the three-point function. Unlike the highly gaussian spectrum of single-field slow-roll inflation, the rapidly-varying equation of state
in our case results in a three-point amplitude that is strongly scale dependent and peaks on small scales. For the simplest potential~(\ref{potent}), in particular, non-gaussianities
are of ${\cal O}(1)$ on the largest scales and grow as $k^2$, resulting in a breakdown of perturbation theory on small scales. Moreover, loop 
corrections dominate over the classical computation on small scales, indicating strong coupling.  

However, these pathologies all result from maintaining the transition phase with large $c$ longer than necessary. Strong coupling can be circumvented by modifying the potential
such that the exponent decreases smoothly from $c$ to $b\ll c$ after the transition phase has already generated an acceptable range of scale invariant fluctuations. We find that as a result
the power spectrum acquires a strong red tilt on small scales and the two-point amplitude is exponentially small. Suppressing the small-scale amplitude
in this way enables perturbation theory to be valid on all scales, both classically and quantum mechanically. By the same token, this modification also addresses all criticisms raised by~\cite{mukhanovlinde}.  Satisfying all requirements comes at a cost, though: the range of nearly scale invariant and gaussian modes is now limited, spanning at most a factor of $10^5$ in $k$ space, or a dozen e-folds, which is sufficient
to account for microwave background and large scale structure observations.

The paper is organized as follows. We begin in Sec.~\ref{std} by reviewing the background dynamics for the adiabatic ekpyrotic mechanism. In Sec.~\ref{2pt}, we calculate
the two-point function, confirming that the power spectrum is scale invariant, and briefly discuss the scalar spectral index and the tensor spectrum. In Sec.~\ref{attract}, we establish that the background
is a dynamical attractor by showing that various physical observables become smaller in time and approach the background solution. Section~\ref{NG} focuses on non-linearities and higher-order correlation functions. We explicitly compute the three-point amplitude for the fiducial potential~(\ref{potent}) and find that it is strongly scale dependent (Sec.~\ref{3ptcalc}). Although most contributions can be well approximated by the horizon-crossing approximation, surprisingly this method fails for the dominant contribution, which instead peaks at late times (Sec.~\ref{horcrossvslw}). This growth in non-linearities results in a breakdown of classical perturbation theory (Sec.~\ref{classicaldiag}) and quantum strong coupling (Sec.~\ref{quantumdiag}). In Sec.~\ref{smoothmodel} we show how these problems can be avoided by modifying the potential as mentioned above, derive various parameter constraints to ensure that non-gaussianities and quantum effects are under control, and offer a few working examples.
We briefly review our main results and discuss prospects for future directions in Sec.~\ref{conclu}.

\section{Background dynamics}
\label{std}

The adiabatic mechanism is most simply realized with the lifted exponential potential~(\ref{potent}), where $V_0> 0$ and $c\gg 1$. This potential is approximately constant and positive at large positive $\phi$, and tends to a negative exponential form for large negative $\phi$. The example is not designed to represent a complete cosmological model; rather, we focus only on a particular range of $\phi$  to illustrate the basic idea behind the adiabatic mechanism.  The regime of interest is the transition when the equation of state rises rapidly from $\epsilon \ll 1$, where the constant term dominates, to $\epsilon \approx c^2/2\gg 1$, where the negative exponential term dominates. We refer to this as the {\it transition} phase.  Similar behavior over this narrow range of $\Delta \phi \ll M_{\rm Pl}$ can be obtained for a wide range of potential functions $V(\phi)$;  we will see that this freedom enables ways of avoiding problems encountered if this first example is considered for all $\phi$.

The form for $V(\phi)$ reminds one of inflationary examples, but our mechanism is emphatically not inflationary in nature.
This can be seen in different ways: $i)$ the universe is slowly {\it contracting} just prior to the onset of the adiabatic mechanism; $ii)$ scale invariant perturbations are generated within one Hubble time, hence the universe is essentially static in the process; and $iii)$ because $\epsilon$ is changing rapidly during the transition, our background evolution violates
the usual slow-roll condition $\eta \ll 1$ of inflation.

Because the scalar field is falling off a steep potential, the evolution is insensitive to the slowly-contracting
background and is, therefore, driven by the potential:
\begin{equation}
\ddot{\phi} \approx -V_{,\phi} = -\frac{c}{M_{\rm Pl}}V_0e^{-c\phi/M_{\rm Pl}}\,.
\label{phieom}
\end{equation}
In particular, the evolution of $\phi$ is oblivious to the constant term $V_0$. As a consistency check, we will see shortly that the transition phase
occurs in less than one Hubble time. Assuming negligible initial kinetic energy, the solution is of the standard ekpyrotic form~\cite{oldek,seiberg,oldek2}
\be
\phi(t) \approx \frac{2M_{\rm Pl}}{c}\log\left(\sqrt{\frac{V_0}{2M_{\rm Pl}^2}}c |t|\right)\,.
\label{phisol}
\ee
(Here $-\infty < t < 0$, with $t=0$ corresponding to $\phi\rightarrow -\infty$.)

The cosmological background, meanwhile, can be inferred from the $\dot{H}$ equation,
\be
\dot{H} = -\frac{1}{2M_{\rm Pl}^2}\dot{\phi}^2 = -\frac{2}{c^2t^2}\,, 
\label{dotH}
\ee
with solution 
\be
H_{\rm tran}(t) = H_0 + \frac{2}{c^2t}\,.
\label{H}
\ee
At sufficiently early times, $H$ is nearly constant, with the constant $H_0$ fixed by the Friedmann equation: $3H^2M_{\rm Pl}^2 = \dot{\phi}^2/2 + V(\phi) \approx V_0$. During the transition phase, the scalar kinetic energy nearly cancels the negative exponential term in the potential, leaving the constant term $V_0$ as the dominant contribution to $H$:
\begin{equation} 
H_0 = -\sqrt{\frac{V_0}{3M_{\rm Pl}^2}}\,.
\end{equation}
As the universe contracts and $|t|$ decreases, eventually the constant term no longer dominates, and 
the solution approaches a standard ekpyrotic phase with $H(t)\approx 2/c^2t$. The end of the transition
phase --- and the onset of the ekpyrotic scaling phase --- occurs when the time-dependent and constant terms become comparable, that is, at 
\begin{equation}
t_{\rm end-tran} \equiv t_{\rm beg-ek} = -\frac{2}{c^2}\frac{1}{|H_0|}\,.
\label{tend}
\end{equation}

The transition phase is also finite in the past. The above derivation neglects gravity, which is a poor approximation
for sufficiently large positive $\phi$ where the potential is flat and Hubble damping is important.
Specifically, the approximation $H\dot{\phi} \ll c V_0e^{-c\phi/M_{\rm Pl}}/M_{\rm Pl}$ implicit in~(\ref{phieom}) is consistent as long as
\begin{equation}
t > t_{\rm beg-tran} \equiv - \frac{1}{|H_0|}\,.
\label{tbegin}
\end{equation}
The transition phase, defined by $t_{\rm beg-tran} < t < t_{\rm end-tran}$, therefore lasts less than a Hubble time, as claimed earlier.
The scale factor, 
\be
a_{\rm tran}(t) \approx 1 + H_0t   + \frac{2}{c^2} \log \left(H_0t \right)\,,
\label{a(t)}
\ee
is approximately constant throughout, and the universe is nearly static. (In integrating~(\ref{H}) to solve for $a(t)$, we have chosen the integration constant such that the log term vanishes at $t_{\rm beg-tran}$.)

Given~(\ref{dotH}),~(\ref{H}) and~(\ref{tend}), the equation of state parameter can be expressed as
\be
\epsilon = -\frac{\dot{H}}{H^2} = \frac{2}{c^2H_0^2}\frac{1}{(t+t_{\rm end-tran})^2}\,.
\label{epsapprox}
\ee
Deep in the transition phase, $|t|\gg |t_{\rm end-tran}|$, the equation of state is rapidly varying, $\epsilon\sim 1/t^2$ --- 
the key to generating a scale invariant spectrum of curvature perturbations. For $|t| \ll |t_{\rm end-tran}|$, meanwhile, the equation of state tends to a large constant,
$\epsilon\rightarrow c^2/2\gg 1$, consistent with an ekpyrotic scaling phase~\cite{oldek,oldek2}:
\begin{equation}
a_{\rm scaling} (t) \sim (-t)^{2/c^2}\;;\qquad H_{\rm scaling}(t) \approx  \frac{2}{c^2t} \,.
\label{ekstd}
\end{equation}
Over the course of the transition phase, therefore, $\epsilon$ grows by a large factor: from $ {\cal O}(1/c^2)$ to ${\cal O}(c^2)$. 

The adiabatic mechanism relies on an exponentially growing
$\epsilon(t)$ and nearly constant $a(t)$, the exact opposite of the exponentially growing $a(t)$ and nearly constant $\epsilon$ characteristic of inflationary cosmology.  In particular, the rate of change of $\epsilon$ is never small:
\be
\eta = \frac{1}{H}\frac{{\rm d}\ln \epsilon}{{\rm d}t} = -c^2 \frac{t_{\rm end-tran}t}{(t+t_{\rm end-tran})^2}
\label{eta}
\ee
ranges from ${\cal O}(1)$ to ${\cal O}(c^2)\gg 1$ during the transition, so the usual slow-roll condition $\eta\ll 1$ is violated throughout.

\section{Power Spectrum}
\label{2pt}

The generation of perturbations is conveniently described by $\zeta$~\cite{zeta1,zeta2}, the curvature perturbation in comoving gauge, $\delta\phi = 0$, $h_{ij} = a^2(t)(1+2\zeta)\delta_{ij}$,
which completely fixes the gauge. The quadratic action governing $\zeta$ is
\be
S_2 = \frac{M_{\rm Pl}^2}{2}\int {\rm d}^3x{\rm d}\tau \;z^2\left[ \zeta'^2 - (\vec{\nabla}\zeta)^2\right]\,,
 \label{actionzeta}
\ee
where primes denote derivatives with respect to conformal time $\tau$, and $z$ is defined as usual by
\begin{equation}
z \equiv a(\tau)  \sqrt{2 \epsilon(\tau)}\,.
\end{equation}
It is convenient to work in terms of the canonically-normalized variable, $v = z\, \zeta$, whose mode functions 
 $v_k$ satisfy
\begin{equation}
v''_k + \left(k^2 - \frac{z''}{z}\right) v_k =0\,.
\label{vpert}
\end{equation}

In the ``transition phase", the scale factor is nearly constant, $a(\tau)\approx 1$, --- this is the slowly contracting background typical of ekpyrotic cosmology ---
hence conformal time and cosmological time are approximately the same: $t\approx \tau$. It follows that $z \approx \sqrt{2\epsilon(t)}$, with
$\epsilon(t)$ given by~(\ref{epsapprox}), and therefore 
\be
\ddot{v}_k + \left(k^2 - \frac{2}{(t+t_{\rm end-tran})^2}\right)v_k = 0\,.
\label{vpertexact}
\ee
This equation is valid throughout the transition phase and deep in the ekpyrotic scaling phase. The approximation $a(t)\approx 1$
eventually breaks down as $t\rightarrow 0^{-}$, since $a(t) \sim (-t)^{2/c^2}$, $\epsilon = c^2/2$, and hence $\ddot{z}/z = 2/c^2t^2$ in the ekpyrotic scaling phase. 
Comparison with $\ddot{z}/z = 2/t_{\rm end-tran}^2$ implied by the late-time limit of~(\ref{vpertexact}) shows that our mode function equation breaks down at $t \sim t_{\rm end-tran}/c$.
But this is well after all modes of interest have exited the Hubble horizon. Therefore,~(\ref{vpertexact}) accurately describes the generation and freeze-out
of scale invariant modes during the transition phase, as well as their Hubble exit during the ekpyrotic scaling phase.

For $|t|\gg |t_{\rm end-tran}|$,~(\ref{vpertexact}) reduces to the same mode function equation as in inflation, 
where $\epsilon$ is nearly constant and $a(t)$ grows exponentially! The two-point function thus generated is
therefore identical to inflation and, in particular, is scale invariant. Indeed, assuming the usual adiabatic vacuum, the solution for the mode functions is
\be
v_k(t) = \frac{e^{-ikt}}{\sqrt{2k}M_{\rm Pl}}\left(1-\frac{i}{k(t+t_{\rm end-tran})}\right)\,.
\ee
Translating back to the curvature perturbation, $\zeta_k=v_k/z$, we obtain
\be
k^{3/2}\zeta_k = \frac{i c |H_0|}{2\sqrt{2}M_{\rm Pl}} \left[1 + i k (t+t_{\rm end-tran})\right]e^{-i k t} \,.
\label{u}
\ee
The corresponding power spectrum as $t\rightarrow 0$, defined by $\langle \zeta_{\vec{k}}\zeta_{\vec{k}'}\rangle = (2\pi)^3\delta^3(\vec{k}+\vec{k}')2\pi^2k^{-3}P_{\zeta}(k)$, is
\be
P_\zeta(k) = \frac{c^2H_0^2}{16\pi^2M_{\rm Pl}^2}\left( 1 + k^2t_{\rm end-tran}^2\right)\,.
\label{zetafin}
\ee
The spectrum is therefore nearly scale invariant for $k|t_{\rm end-tran}|\ll 1$, corresponding to modes that freeze out during the transition phase. For $k|t_{\rm end-tran}|\gg 1$, 
corresponding to modes that freeze out during the ekpyrotic scaling phase, the spectrum is far from scale invariant, $P_\zeta\sim k^2$, consistent with
the strong blue tilt for $\zeta$ of the original ekpyrotic generation mechanism~\cite{oldek2,gratton,robertek,nicolis}. (The blue tilt can be modified by choosing a different $V(\phi)$, as described below.)
The range of scale invariant modes is determined by the duration of the transition phase:
\be
\frac{k_{\rm max}}{k_{\rm min}} = \frac{t_{\rm beg-tran}}{t_{\rm end-tran}} = \frac{c^2}{2}\,.
\ee
As we will see shortly, requiring that the scale invariant range overlaps with the largest observable scales today and that the magnitude of $\zeta$ matches observations
force $c$ to be exponentially large.

\begin{figure} %  figure placement: here, top, bottom, or page
   \centering
   \includegraphics[width=5.0in]{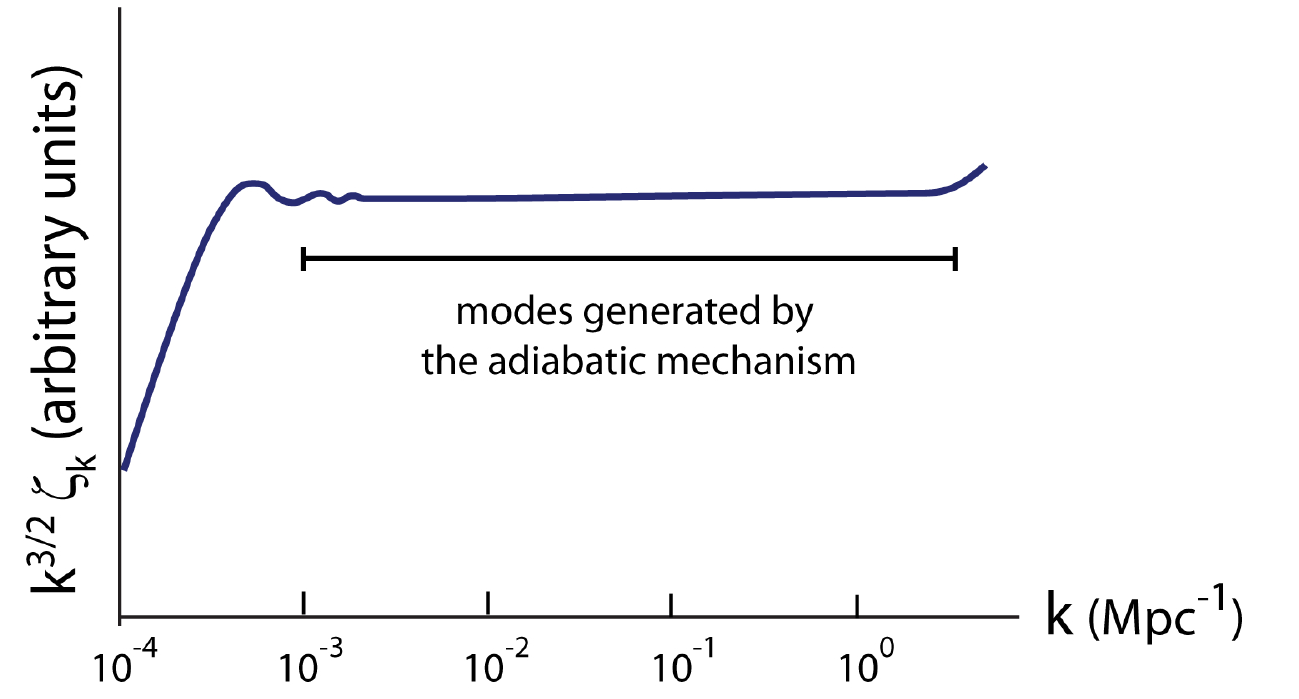}
   \caption{Numerical computation of the perturbation amplitude $k^{3/2} \zeta$ vs. $k$ generated by the adiabatic mechanism.  The behavior of modes with larger and smaller $k$ depends on the larger scenario in which the mechanism is embedded and beyond the consideration of this paper. The range of scale invariant modes is in good agreement with the analytical treatment.}
   \label{zetak}
\end{figure}

Our analytical treatment is borne out by numerical analysis. Using $z = c(-t)^{2/c^2}/(1+c^2H_0t/2)$ to cover the transition and scaling phases, we
integrate~(\ref{vpert}) with $c=140$ and $|H_0|=  10^{-3}$ over the interval $-5\times |H_0|^{-1} < t <  -10^{-9} \times |H_0|^{-1}$, over the range of modes
$10^{-2}\times |H_0| < k < 10^4 \times |H_0|$. Figure~\ref{zetak} shows the resulting spectrum. 
The shortest-wavelength modes are barely outside the Hubble radius by the end of the integration,
which occurs deep in the ekpyrotic scaling phase. Modes with $k \;\lsim\; |H_0| = 10^{-3}$ begin outside the $\zeta$-horizon
at the initial time and hence are not scale invariant. The numerical results show good agreement with the expected range
$10^{-3} \;\lsim\; k  \;\lsim\; 10$ of scale invariant modes. 

The mode function solution~(\ref{u}) clearly tends to a constant at late times, $\zeta \rightarrow {\rm constant}$ as $t\rightarrow 0$, again as in inflation. Since $\zeta$ represents a perturbation of the scale factor in this limit~\cite{separateuniverse}, $\zeta \approx \delta a/a$, this implies that the ``transition phase" evolution is a dynamical attractor. This statement will be made more precise in Sec.~\ref{attract}; but, for the moment, let us contrast this with a contracting, dust-dominated universe, corresponding to $a(\tau)\sim (-\tau)^2$ and $\epsilon = 3/2$. This background has often been hailed as the dual to the inflationary mechanism~\cite{gratton,dust,latham} since $z''/z = 2/\tau^2$ in this case, exactly as in inflation. But because $z \sim (-\tau)^2$, the curvature perturbation grows at late times, $\zeta \sim 1/(-\tau)^3$, indicating that the background is unstable. By contrast, the spectrum generated by a slowly-contracting universe with rapidly-varying equation of state, as proposed here, has identical two-point function and long-wavelength evolution for $\zeta$ as inflationary cosmology.

\subsection{Observational Constraints}
\label{cons2}

Our power spectrum must meet two observational criteria. Firstly, the amplitude of the power spectrum over the scale
invariant range ($k|t_{\rm end-tran}|\ll1$) must match the observed value
\be
\frac{c^2H_0^2}{16\pi^2M_{\rm Pl}^2} \sim 10^{-10}\,,
\label{ampcons}
\ee
which, given $H_0$, fixes $c$. Secondly, the scale invariant range must overlap with the scales probed by cosmic
microwave background and large-scale structure observations, which requires that the comoving scale $1/|H_0|$
encompass the entire observable universe. Assuming that the magnitude of $H$ at the
end of the ekpyrotic phase, $|H_{\rm ek-end}|$, is comparable to that at the onset of the expanding, radiation-dominated phase, 
then we must demand 
\be
\frac{|H_{\rm ek-end}|}{|H_0|} \;\gsim\; \frac{a_{\rm ek-end}|H_{\rm ek-end}|}{a_{\rm today}H_{\rm today}} \sim \sqrt{\frac{|H_{\rm ek-end}|}{H_{\rm today}}}\,,
\ee
where in the last step we have assumed a radiation-dominated evolution until the present epoch, for simplicity. In other words,
\begin{equation}
|H_0| \;\lsim \; \sqrt{|H_{\rm ek-end}|H_{\rm today}} \approx 10^{-30} \sqrt{|H_{\rm ek-end}| M_{\rm Pl}}\,,
\label{V0cons}
\end{equation}
which constrains $H_0$ in terms of the reheating scale. For Grand-Unified (GUT) reheating scale, $|H_{\rm ek-end}| \sim 10^{12}$~GeV, the above condition is satisfied for $|H_0| \sim 10^{-3}$~meV, corresponding to $V_0\sim (10\;{\rm GeV})^4$. The constraint~(\ref{ampcons}) on the power spectrum amplitude then fixes $c = 10^{28}$. For electroweak (EW) reheating,
$|H_{\rm ek-end}| \sim {\rm meV}$, we similarly get $V_0 \sim (10^{-2}\;{\rm MeV})^4$ and $c = 10^{40}$.

Thus our mechanism requires exponentially large values of $c$. To recap, this is because 
the universe is nearly static during the generation of perturbations, hence the Hubble parameter must be
small relative to $M_{\rm Pl}$ for perturbations to overlap with the scales probed by observations. This in turn requires a very steep exponential potential
in order for the amplitude to match observations. Although we have focused on pure exponential potentials, for simplicity, the exponentially large values of
 $c$ required could be achieved, for instance, in the Conlon-Quevedo potential~\cite{CQ}, $V(\phi) \sim \exp(-\gamma\phi^{4/3})$, for  large $\phi$.

\subsection{Other Observables}

To conclude this Section, we briefly comment on two other observables, namely the scalar spectral index and the tensor spectrum.
Since the values of $c$ of interest are exponentially large, the departures from scale invariance are unobservable for the potential considered thus far. However, in this overly simple example, the ekpyrotic phase never ends and the universe does not bounce.  In a complete model, the exponent $c$ is replaced by $c(\phi)$ which is exponentially large during most of the ekpyrotic phase but is made to fall below one at some given $\phi$ in order to end the ekpyrotic phase.    The variation in $c(\phi)$ results a small red tilt, as favored by observations~\cite{wmap7}:
\begin{equation}
V(\phi) = V_0 \left(1- e^{-\int {\rm d}\phi \; c(\phi)/M_{\rm Pl}}\right)\,.
\label{tiltpot}
\end{equation}
As shown in~\cite{PRL}, the resulting tilt is
\begin{equation}
n_s - 1  = -4M_{\rm Pl}\frac{c_{,\phi}}{c^2}\,.
\end{equation}
Since $\phi$ is decreasing in our solution, the spectral tilt will be slightly red if $c_{,\phi}> 0$. For instance, if $c(\phi)$ changes smoothly by ${\cal O}(c)$ during the transition, then 
\begin{equation}
n_s - 1 \approx \frac{4M_{\rm Pl}}{c\Delta\phi} \approx \frac{2}{\log(t_{\rm end-tran}/t_{\rm beg-tran})} = -\frac{2}{\log(c^2/2)}\,,
\end{equation}
which gives $n_s \approx 0.98$ for $10^{40} > c  > 10^{28}$,  ranging from EW to GUT-scale reheating. Allowing for various ${\cal O}(1)$ factors that were dropped in this estimate,
the generic answer is $1-n_s  \approx {\rm few} \;\%$, in good agreement with observations of the cosmic microwave background~\cite{wmap7}.

Gravitational waves, meanwhile, are not appreciably excited because the background is slowly evolving. Tensor perturbations maintain their adiabatic vacuum normalization,
$h_k \sim 1/\sqrt{k}$, resulting in a strong blue tilt for their spectrum
\be
P_h(k) \sim k^3|h_k|^2 \sim k^2\,,
\ee
corresponding to $n_{\rm T} = 2$. As in earlier renditions of ekpyrotic cosmology~\cite{oldek,lathamgw}, the primordial tensor amplitude is therefore exponentially suppressed on the largest scales. The dominant gravitational wave background at long wavelengths is the secondary gravitational waves induced by the energy density fluctuations, roughly $10^{-5}$ times smaller than the primordial fluctuations~\cite{second}.  Detection
of primordial gravitational waves, for instance through cosmic microwave background B-mode polarization, would unequivocally rule out this mechanism.

\section{Stability}
\label{attract}

The fact that $\zeta\rightarrow {\rm constant}$ at long wavelengths suggests that our background is a dynamical attractor~\cite{weinbergzeta}.
To make this statement precise, we will show below that all physical observables become smaller in time and
approach the background solution. We focus on the transition phase, since the ekpyrotic scaling phase has already been
shown to be an attractor~\cite{gratton,erick,nicolis,pretorius}.

Following~\cite{nicolis}, we find that synchronous gauge, in which $g_{00} =-1$ and $g_{0i} = 0$, is a well-suited coordinate system to study stability.
The scalar perturbations in this gauge are encoded in the spatial components of the metric,
\be
g_{ij} = a^2\left\{\left(1+2\zeta- 2H\int^{t}{\rm d}t'\frac{\dot{\zeta}}{H}  \right)\delta_{ij}
+2\partial_i\partial_j \int^{t} {\rm d}t' \left[ -\frac{\epsilon}{\vec{\nabla}^2}\dot{\zeta} + \frac{\zeta}{a^2H} - \frac{1}{a^2}\int^{t'}{\rm d}t''\frac{\dot{\zeta}}{H} \right] \right\}\,,
\label{gij}
\ee
as well as in scalar field fluctuations:
\be
\delta\phi = -\dot{\phi} \int^{t}{\rm d}t'\frac{\dot{\zeta}}{H} \,.
\label{delphisync}
\ee

We first show that all metric components tend to their unperturbed value, up to rescaling of coordinates, while scalar field perturbations tend to zero.
First note that the growing mode solution to~(\ref{vpert}) has the following long-wavelength expansion
\be
\zeta_k = \zeta_0\left(1+ \frac{1}{2}k^2t^2 + {\cal O}(k^3t^3)\right)\,,
\label{zetasync}
\ee
where the time-dependent amplitude $\zeta_0$ is determined by initial conditions. For the matter perturbation $\delta\phi$, this implies
\be
\delta\phi \approx \frac{M_{\rm Pl}}{c|H_0|}k^2\zeta_0 t \,,
\label{delphilong}
\ee
which becomes increasingly small in time. For the metric, the coefficient of the $\delta_{ij}$ term in~(\ref{gij}) gives
\be
1+2\zeta- 2H\int^{t}{\rm d}t'\frac{\dot{\zeta}}{H} \approx 1 + 2\zeta_0 + \ldots\,,
\ee
where the ellipses indicate terms that become negligible in time. Thus this term goes to a constant, which can be absorbed in a (time-independent) spatial gauge transformation.

Furthermore, since $\epsilon \approx 6M_{\rm Pl}^2/c^2V_0t^2$ during the transition phase, the $\partial_i\partial_j$ term in (\ref{gij}) becomes
\be
2\partial_i\partial_j \int^{t} {\rm d}t' \left[ -\frac{\epsilon}{\vec{\nabla}^2}\dot{\zeta} + \ldots\right]
=  - \frac{2k_ik_j\zeta_0}{H_0^2}\cdot \frac{2}{c^2}\log \left( H_0t\right)  + \ldots\,,
\label{didjterm}
\ee
where a suitable spatial diffeomorphism has been assumed to normalize the log. The growth of this log term looks at first sight dangerous, but note
that its time-dependence exactly matches the log term in $a(t)$ --- see~(\ref{a(t)}). This contribution therefore renormalizes the departure from de Sitter space of the background solution, and, as such, does not signal an instability. 

It is also instructive to study the stability of curvature invariants. Starting with the Ricci curvature of the 3-metric, only the $\delta_{ij}$ term contributes to this quantity since the $\partial_i\partial_j$ term is pure gauge:
\be
R^{(3)} \sim \frac{k^2}{a^2}\left(1 + 2\zeta- 2H\int^{t}{\rm d}t'\frac{\dot{\zeta}}{H}  \right)\rightarrow \frac{k^2}{a^2} \zeta_0\,.
\label{R3}
\ee
The 3-curvature thus goes to a constant, which is acceptable. (The amplification to a constant $R^{(3)}$ is precisely how we generate scale invariant perturbations.)
Similarly, the perturbation in the extrinsic curvature tensor, $K_{ij} = \dot{g}_{ij}/2$, is of order
\be
\frac{\delta K}{\bar{K}} \sim  \frac{k^2\zeta_0}{H_0^2} + \ldots\,,
\label{delK}
\ee
where we have used $\bar{K} =3H(t) = 3(H_0 + 2/c^2t)$. Thus, the perturbation in $K$ thus also tends to a constant at late times.

These results, together with earlier analyses of standard ekpyrotic scenarios~\cite{gratton,erick,nicolis,pretorius}, establish that our cosmological background, consisting of a transition phase followed
by an ekpyrotic scaling phase, is an attractor solution. The breadth of its basin of attraction is a question that requires numerical analysis and will be discussed elsewhere~\cite{austin}. 
The fact that small scale modes are highly non-gaussian and that non-linearities grow after freeze-out, as discussed in the next Section, suggests that the basin of attraction is limited to small perturbations. Moreover, we will see in Sec.~\ref{smoothmodel} that our mechanism can produce at most a dozen e-folds of nearly scale invariant and gaussian modes.
This range, while sufficiently broad to account for large scale observations, does not leave much room to wash out arbitrary initial conditions. 
Note that we cannot draw firm conclusions about the evolution {\it before} the transition phase, as this is clearly model dependent.
If one insists on trusting the lifted exponential potential~(\ref{potent}) at large $\phi$, then the universe is initially in a {\it contracting} de Sitter phase,
which is of course unstable to kinetic domination. But, as mentioned before, there is considerable freedom in specifying the potential in the pre-transition phase. In~\cite{austin}, for instance,
we present a version of the scenario for which the background solution is initially expanding and therefore stable for all times.

\section{Non-Gaussianities and Strong Coupling}
\label{NG}

While a phase of rapidly-varying $\epsilon(t)$ yields an identical power spectrum as inflation, we will see that
the degeneracy is broken at the three-point level. Non-gaussianities are strongly scale dependent, with the dominant contribution
growing as $k^2$. This implies that our mechanism can only generate a finite range of modes within the perturbative regime.
A related pathology, also discussed below, is that the theory becomes strongly coupled on small scales, invalidating
the classical description for these modes. 

These problems all have a common origin: the transition phase with large $c$ is maintained longer than needed --- a consequence of the simple $V(\phi)$ considered so far. In Sec.~\ref{smoothmodel}, we will consider altering
the pure exponential potential so as to terminate the transition phase before the modes with unacceptably large non-linearities are generated, thereby shutting
off power on small scales. This achieves the desired goal of avoiding strong coupling and large non-gaussianities, while providing a range of scale invariant modes on observational scales sufficient to account for microwave background and large scale structure measurements.

\subsection{Computing the three-point Amplitude}
\label{3ptcalc}

For a canonical scalar field with unit sound speed, the exact action to cubic order in $\zeta$ is given by~\cite{maldacena,seery,kachru}
\begin{eqnarray} 
\label{action3}
S_{3}&=&\int {\rm d}t {\rm d}^3x \left\{
a^3\epsilon^2\zeta\dot{\zeta}^2 + a\epsilon^2\zeta(\vec{\nabla}\zeta)^2-
2a \epsilon\dot{\zeta}\vec{\nabla}
\zeta\cdot \vec{\nabla} \chi \right.\nonumber \\ &+& \left.
\frac{a^3\epsilon}{2}\dot{\eta}\zeta^2\dot{\zeta}
+\frac{\epsilon}{2a}\vec{\nabla}\zeta\cdot\vec{\nabla}
\chi \nabla^2 \chi +\frac{\epsilon}{4a}\nabla^2\zeta(\vec{\nabla}
\chi)^2+ 2 f(\zeta)\left.\frac{\delta L_{(2)}}{\delta \zeta}\right\vert_1 \right\} \,,
\end{eqnarray}
where spatial derivatives are contracted with the Euclidean metric $\delta_{ij}$, 
and $\chi$ is defined as
\begin{equation}
\nabla^2 \chi = a^2 \epsilon\dot{\zeta}\,.
\label{chidef}
\end{equation}
The last term, proportional to the linearized equations of motion, 
\begin{eqnarray}
\left.\frac{\delta
L_{(2)}}{\delta\zeta}\right\vert_1 &=& a
\left( \frac{{\rm d}\nabla^2\chi}{{\rm d}t}+H\nabla^2\chi
-\epsilon\nabla^2\zeta \right)\,,
\end{eqnarray}
can be absorbed as usual through a field redefinition 
\begin{eqnarray}
\zeta \rightarrow \zeta+f(\zeta)\,,
\label{redef}
\end{eqnarray}
where
\begin{eqnarray} \label{redefinition}
f(\zeta)&=&\frac{\eta}{4}\zeta^2+\frac{1}{H}\zeta\dot{\zeta}+
\frac{1}{4a^2H^2}[-(\vec{\nabla}\zeta)^2+\nabla^{-2}(\nabla^i\nabla^j(\nabla_i\zeta\nabla_j\zeta))] \nonumber \\
&+&
\frac{1}{2a^2H}[\vec{\nabla}\zeta\cdot \vec{\nabla}\chi-\nabla^{-2}(\nabla^i\nabla^j(\nabla_i\zeta\nabla_j\chi))] \,.
\end{eqnarray}

At first order in perturbation theory and in the interaction picture, the three-point function is
\begin{equation} \label{interaction}
\langle
\zeta(t,\textbf{k}_1)\zeta(t,\textbf{k}_2)\zeta(t,\textbf{k}_3)\rangle=
-i\int^{0}_{-\infty}{\rm d}t^{\prime}\langle[
\zeta(t,\textbf{k}_1)\zeta(t,\textbf{k}_2)\zeta(t,\textbf{k}_3),H_{\rm int}(t^{\prime})]\rangle \,,
\end{equation}
where $H_{\rm int} = - L_{3}$, up to interactions that are higher-order in the number of fields. 
As usual we expand the quantum field $\zeta$ in terms of creators and annihilators,
\begin{equation} 
\label{modes} 
\zeta(\vec{k},t) = \zeta_k(t)a(\vec{k}) + \zeta_k^*(t) a^\dagger(-\vec{k}) \,,
\end{equation}
with commutation relations $[a(\kk), a^\dagger(\kk')] = (2 \pi)^3 \delta^3(\kk - \kk')$.
Although the upper limit of integration in~(\ref{interaction}) has been set at $t=0$, strictly speaking our approximation $a(t)\approx 1$
assumed throughout breaks down at a time $t \sim t_{\rm end}/c$, as discussed in Sec.~\ref{2pt}. We have checked that for the modes of interest this makes little difference to the final answer,
hence we are justified in integrating all the way to $t=0$ setting $a(t) = 1$. To simplify the expressions we set $M_{\rm Pl} = 1$ for the remainder of this section.   

The three-point function receives contributions from each interaction term in~(\ref{action3}). The dominant contributors, it turns out, are the last two terms in the cubic action,
both which are of order $\epsilon^3$. The next-to-leading contribution is the $\dot{\eta}$ term. We present explicit calculations of these contributions and refer the 
reader to the Appendix for the rest of the calculation. 

\begin{itemize}

\item{\bf The $\epsilon^3$ contributions}: The $\epsilon^3$ terms give the combined interaction Hamiltonian
\be
H_{\rm int} =  -\frac{\epsilon^3}{4}\int {\rm d}^3x\left(\nabla^2\zeta\frac{\vec{\nabla}}{\nabla^2}\dot{\zeta}\frac{\vec{\nabla}}{\nabla^2}\dot{\zeta}+2\dot{\zeta}\vec{\nabla}\zeta \frac{\vec{\nabla}}{\nabla^2}\dot{\zeta}\right)\,.
\ee
Applying the canonical commutation relations, the three-point correlation function~(\ref{interaction}) in this case reduces to
\begin{multline}
\langle \zeta(\textbf{k}_1)\zeta(\textbf{k}_2)\zeta(\textbf{k}_3)\rangle_{\epsilon^3}\, =\,
i (2 \pi)^3 \delta^3(\kk_1+\kk_2+\kk_3) \zeta_{k_1}(0)\zeta_{k_2}(0)\zeta_{k_3}(0) \\ \times 
\int_{-\infty+i\varepsilon}^{0} {\rm d} t \, \frac{\epsilon^3}{4} \left(\frac{k_1^2}{k_2^2}\zeta_{k_1}^*(t)\,\frac{{\rm d} \zeta_{k_2}^*(t)}{{\rm d} t} + 2\frac{{\rm d} \zeta_{k_1}^*(t)}{{\rm d} t}\zeta_{k_2}^*(t) \right)\frac{\vec{k}_2\cdot \vec{k}_3}{k_3^2}\frac{{\rm d} \zeta_{k_3}^*(t)}{{\rm d} t} + {\rm perm.} + {\rm c.c.} \,,
\label{eps33pt}
\end{multline}
where the small imaginary part at $t\rightarrow -\infty$ projects onto the adiabatic vacuum state. Substituting the mode functions~(\ref{u}) and
using~(\ref{epsapprox}) for $\epsilon(t)$, we obtain
\bea
\nonumber
& & \langle \zeta(\textbf{k}_1)\zeta(\textbf{k}_2)\zeta(\textbf{k}_3)\rangle_{\epsilon^3} = -\frac{1}{128}(2\pi)^3\delta^3(\kk_1+\kk_2+\kk_3) K\left(\sum_ik_i^3 - \sum_{i\neq j}k_ik_j^2+2k_1k_2k_3\right)  \\
& & \;\;\;\;\;\;\;\;\;\;\;\;\;\;\;\;\;\;\;\;\;\;\;\; \times  \; {\rm Im} \left\{\prod_j (1+ik_jt_{\rm end-tran}) \int_{-\infty+i\varepsilon}^{0} {\rm d}t \, \frac{3-iK(t+t_{\rm end-tran})}{(t+t_{\rm end-tran})^4}e^{iKt} \right\}\,,
\eea
where $K\equiv k_1+k_2+k_3$.

As usual it is convenient to express the three-point function by factoring out appropriate powers of the power spectrum and defining an amplitude ${\cal A}$ as follows
\begin{equation}
\langle \zeta(\textbf{k}_1)\zeta(\textbf{k}_2)\zeta(\textbf{k}_3)\rangle = (2\pi)^7 
\delta^3(\kk_1+\kk_2+\kk_3) P_\zeta^{\;2} \frac{{\cal A}}{\prod_j k_j^3}\,,
\label{Adef}
\end{equation}
where $P_\zeta$ is given by~(\ref{zetafin}). Fortunately, the integrand is a total derivative:
\be
\int_{-\infty+i\varepsilon}^{0} {\rm d}t \, \frac{3-iK(t+t_{\rm end-tran})}{(t+t_{\rm end-tran})^4}e^{iKt} = - \int_{-\infty+i\varepsilon}^{0} {\rm d}t \, \frac{{\rm d}}{{\rm d}t}\left(\frac{e^{iKt}}{(t+t_{\rm end-tran})^3}\right) = -\frac{1}{t_{\rm end-tran}^3}\,.
\ee
Substituting~(\ref{tend}) for $t_{\rm end-tran}$ and focusing on the long wavelength limit $K|t_{\rm end-tran}|\ll 1$, which is appropriate for the modes of interest, the three-point amplitude is
\be
{\cal A}_{\epsilon^3} = \frac{K^2}{32H_0^2}\left(\sum_ik_i^3 - \sum_{i\neq j}k_ik_j^2+2k_1k_2k_3\right)\,.
\label{eps3}
\ee
Thus this scales as $K^2/H_0^2$ and, as we will see, dominates over all other contributions on scales $K\;\gsim\;|H_0|$. 

\item{\bf The $\dot{\eta}$ contribution}: The interaction Hamiltonian for this contribution is
\be
H_{\rm int} = -\int {\rm d}^3x \,\frac{1}{2}\epsilon\dot{\eta} \zeta^2\dot{\zeta}  \,.
\ee
Substituting~(\ref{epsapprox}) and~(\ref{eta}) for $\epsilon(t)$ and $\eta(t)$, respectively, we obtain the three-point contribution
\bea
\nonumber
& & \langle \zeta(\textbf{k}_1)\zeta(\textbf{k}_2)\zeta(\textbf{k}_3)\rangle_{\dot{\eta}} = 
(2 \pi)^3 \delta^3(\kk_1+\kk_2+\kk_3) \frac{c^4|H_0|^3}{64\cdot \prod_jk_j^3} {\rm Im} \left\{\prod_j (1+ik_jt_{\rm end-tran}) \right.\\
\nonumber
& & \left. \times \int_{-\infty+i\varepsilon}^{0} {\rm d}t \frac{(t-t_{\rm end-tran})e^{iKt}}{(t+t_{\rm end-tran})^4}\left[\sum_ik_i^2-i\sum_{i\neq j}k_ik_j^2(t+t_{\rm end-tran}) - Kk_1k_2k_3(t+t_{\rm end-tran})^2\right]\right\} \\
\label{dotetazeta}
\eea
Performing these various integrals, the corresponding three-point amplitude is given by, in the long wavelength ($K|t_{\rm end-tran}|\ll 1$) limit,
\be
{\cal A}_{\dot{\eta}} = - \frac{\pi}{8}\frac{K}{|H_0|} \left( \frac{K}{2}\sum_ik_i^2 - \sum_{i\neq j}k_ik_j^2 +k_1k_2k_3\right)\,.
\label{Adoteta}
\ee
This contribution scales as $K/|H_0|$ and is therefore subdominant relative to~(\ref{eps3}) on scales $K\;\gsim\;|H_0|$.

\end{itemize}

The remaining contributions, computed in the Appendix, are all suppressed by $1/c^2$ relative to~(\ref{eps3}). The full three-point amplitude can be well approximated by~(\ref{eps3}), at least on scales $K\; \gsim\; |H_0|$:
\be
{\cal A} \approx \frac{K^2}{32H_0^2}\left(\sum_ik_i^3 - \sum_{i\neq j}k_ik_j^2+2k_1k_2k_3\right)\,.
\label{ampfinal}
\ee
As a check, note that this satisfies Maldacena's ``consistency" relation~\cite{maldacena}: in the squeezed limit $k_3 \ll k_1 \approx k_2\equiv k$,
we have ${\cal A}\rightarrow 0$, consistent with our neglecting departures from scale invariance in computing the three-point function. (Although derived in the context of inflation,
Maldacena's relation applies here because our satisfies its two keys assumptions: single field theory and $\zeta\rightarrow {\rm constant}$ as $k\rightarrow 0$~\cite{leonardo}.)
Instead our amplitude peaks for equilateral configurations, $k_i= K/3$. The shape dependence is qualitatively similar to higher-derivative inflationary models~\cite{paoloshape}.

Following standard practice, the three-point amplitude translates into a value for $f_{\rm NL}^{\rm equil.}$, defined at the
equilateral configuration~\cite{paoloshape}:
\be
f_{\rm NL}^{\rm equil.} \equiv 30\frac{{\cal A}_{k_i=K/3}}{K^3}  \approx - \frac{5}{144}\frac{K^2}{H_0^2}\,.
\label{fNLfinal}
\ee
Unlike the power spectrum, the three-point function is thus strongly scale dependent: $f_{\rm NL}^{\rm equil.}$ is $\lsim\; {\cal O}(1)$
on the largest scales ($K \sim |H_0|$) and grows as $K^2$. This is in stark contrast with the small and nearly scale invariant $f_{\rm NL}$ predicted by single-field,
slow-roll inflation. The degeneracy of our mechanism with inflation established at the two-point level is therefore strongly broken at the three-point level.

The strong scale dependence of~(\ref{fNLfinal}) implies that perturbation theory breaks down on relatively large scales. Specifically, the 
perturbative expansion parameter is $f_{\rm NL}\zeta$, with $\zeta\sim 10^{-5}$, hence non-linearities dominate for $K \;\gsim\; 10^{5/2} |H_0|$. 
We will have more to say about this in Sec.~\ref{classicaldiag}. In fact, we will see in Sec.~\ref{quantumdiag} that on even smaller scales
($K\;\gsim\; 10^5|H_0|$) quantum corrections dominate the classical result, indicating
strong coupling. All of these problems have a common origin: the transition phase with large $c$ is maintained longer than needed. A simple solution,
discussed in Sec.~\ref{smoothmodel}, is to alter the pure exponential potential so as to terminate the transition phase before these problem emerge. This restores
perturbative control in two ways: 1. altering the evolution of $\epsilon(t)$ suppresses the dominant $\epsilon^3$ contribution, thereby expanding the range
of perturbative modes; 2. terminating the transition phase suppresses $\zeta$ on smaller scales --- the spectrum tilts strongly to
the red and then flattens out at an exponentially smaller amplitude with an acceptable non-gaussianity ($f_{\rm NL} \zeta \ll 1$) throughout. 
This leaves us with a finite range ($|H_0| \;\lsim \; K \;\lsim\; 10^5|H_0|$) of scale invariant modes, which is sufficient to account for observations.

\subsection{Horizon Crossing vs Long Wavelength Approximations}
\label{horcrossvslw}

A standard, back-of-the-envelope method for estimating $f_{\rm NL}$ is to compare the cubic and quadratic actions for $\zeta$ at freeze-out~\cite{louissarah}:
\be
f_{\rm NL} \sim \zeta^{-1}\left.\frac{S_3}{S_2}\right\vert_{\rm freeze-out}\,.
\label{fNLest}
\ee
This is because non-gaussianities typically peak at horizon-crossing --- deep inside the horizon, modes are approximately in the vacuum state,
whereas far outside the horizon, interactions are suppressed by time derivatives and spatial gradients, which are small relative to Hubble in that regime. 

The situation in our case is trickier: because the various cubic interactions have coefficients such as $\epsilon^2$ or $\epsilon^3$ that {\it grow} rapidly in time,
there is a competition between this growth and the derivative suppression. It turns out that for most of the cubic interactions these two effects nearly balance out,
such that the horizon-crossing approximation provides a good estimate. Consider the $\dot{\eta}$ contribution, for concreteness. Since time and spatial
derivatives are comparable at horizon crossing, we can approximate $\dot{\zeta} \sim \vec{\nabla}\zeta \sim k\zeta$ in evaluating~(\ref{fNLest}) and obtain
\be
f_{\rm NL}^{\;\dot{\eta}} \sim \zeta^{-1}\left.\frac{S_3}{S_2}\right\vert_{k|t| = 1} \sim \zeta^{-1}\left.\frac{\epsilon\, \dot{\eta}\,\zeta^2\dot{\zeta}}{\epsilon\, \dot{\zeta}^2}\right\vert_{k|t| = 1} = \left.\frac{\dot{\eta}\,\zeta}{\dot{\zeta}}\right\vert_{k|t| = 1} \sim \frac{\dot{\eta}}{k} \sim \frac{k}{|H_0|}\,,
\label{fNLestdoteta}
\ee 
where in the last step we have used $\dot{\eta} \simeq 2H_0^{-1}t^{-2} \sim k^2H_0^{-1}$ at horizon crossing. This result agrees with the
parametric dependence of~(\ref{Adoteta}). Similarly for all contributions calculated in the Appendix.

The key exceptions are the $\epsilon^3$ terms. The vertex increases as $\epsilon^3 \sim 1/t^6$ during the transition phase, and this rapid growth overwhelms the derivative suppression.
The three-point contribution therefore peaks at late times, well after the modes of interest have frozen out. Indeed, the horizon-crossing approximation fails miserably in this case:
\be
f_{\rm NL}^{\;\epsilon^3} \sim \zeta^{-1}\left.\frac{S_3}{S_2}\right\vert_{k|t| = 1} \sim \left.\epsilon^2\right\vert_{k|t| = 1} \sim \frac{k^4}{c^4H_0^4}\,,
\ee
which greatly underestimates the exact answer $\sim k^2/H_0^2$. We can shed further light on this contribution by first taking the
long wavelength limit of the mode functions~(\ref{u}). Up to an irrelevant constant phase, the relevant terms are
\be
\zeta_k = C_k \left(1 + \frac{1}{2}k^2y^2 + \frac{i}{3}k^3y^3 + \ldots \right)\,,
\label{zetalw}
\ee
where $y \equiv t + t_{\rm end-tran}$, and
\be
C_k \equiv  \frac{i c |H_0|}{2\sqrt{2}M_{\rm Pl}k^{3/2}} \left[1 + i k t_{\rm end-tran} \right]\,.
\label{Ck}
\ee
By inspection, it turns out that the dominant contribution to~(\ref{eps33pt}) comes from the imaginary part of the integrand:
\bea
\nonumber
\left(\frac{k_1^2}{k_2^2}\zeta_{k_1}^*\,\dot{\zeta}_{k_2}^* + 2\dot{\zeta}_{k_1}^*\zeta_{k_2}^* \right)\frac{\vec{k}_2\cdot \vec{k}_3}{k_3^2}\dot{\zeta}_{k_3}^* + {\rm perm.} &=& 
\left(\prod_j C_{k_j}^\dagger\right) \left\{2i K^2 \left(\sum_ik_i^3 - \sum_{i\neq j}k_ik_j^2+2k_1k_2k_3\right) y^3 \right. \\
& & \;\;\;\;\;\;\;\;\;\;\;\;\;\;\;\;\;\;\;\;\;\;\;\;  + \left. {\rm real\;part} \large\right\}\,.
\label{imeps3}
\eea
As expected, this decreases in time due to the derivative suppression. The trouble is that the $\epsilon^3 \sim 1/y^6$ grows even faster. Indeed, the three-point correlation is
\bea
\nonumber
\langle \zeta(\textbf{k}_1)\zeta(\textbf{k}_2)\zeta(\textbf{k}_3)\rangle_{\epsilon^3} &=&  -\frac{1}{128\prod_jk_j^3}(2\pi)^3\delta^3(\kk_1+\kk_2+\kk_3)K^2\left(\sum_ik_i^3 - \sum_{i\neq j}k_ik_j^2+2k_1k_2k_3\right)   \\
\nonumber
&\times &   \int_{-\infty+i\varepsilon}^{t_{\rm end-tran}} \frac{{\rm d}y}{y^3} + {\rm c.c.} \\
\nonumber
&=& \frac{1}{128\prod_jk_j^3}(2\pi)^3\delta^3(\kk_1+\kk_2+\kk_3) \frac{K^2}{t_{\rm end-tran}^2} \left(\sum_ik_i^3 - \sum_{i\neq j}k_ik_j^2+2k_1k_2k_3\right)\,. \\
\eea 
Substituting the expression~(\ref{tend}) for $t_{\rm end-tran}$, we obtain the amplitude
\be
{\cal A}_{\epsilon^3} = \frac{K^2}{32H_0^2}\left(\sum_ik_i^3 - \sum_{i\neq j}k_ik_j^2+2k_1k_2k_3\right)\,,
\label{eps3lwfirstshot}
\ee
which agrees precisely with~(\ref{eps3}). In other words, because the $y$-integral strongly peaks at $t_{\rm end-tran}$,
the long wavelength approximation reproduces the exact answer. 

\subsection{Classical Perturbation Theory}
\label{classicaldiag}

We have seen that the perturbative expansion parameter $f_{\rm NL}\zeta$ grows larger than unity on small scales.
This breakdown of (classical) perturbation theory can be seen in other observables, such as perturbations in the energy density,
$\delta \rho/\bar{\rho}$, in synchronous gauge. (An equivalent discussion applies to the extrinsic curvature perturbation in this gauge, given by~(\ref{delK}), but here we choose to focus on $\delta \rho/\bar{\rho}$ to parallel the discussion in~\cite{mukhanovlinde}.) With $\bar{\rho} \approx 3H_0^2M_{\rm Pl}^2$ during the transition phase, we have, at linear order, 
\be
\frac{\delta\rho}{\bar{\rho}} = \frac{ \dot{\phi}\delta\dot{\phi} + V_{,\phi}\delta\phi}{H_0^2M_{\rm Pl}^2} \sim \epsilon \frac{k^2t}{H_0}\,\zeta \,,
\label{delrho1}
\ee
where we have used~(\ref{phisol}) and~(\ref{delphilong}). And since $\epsilon \sim 1/t^2$ during the transition phase, this clearly peaks at the end of the transition phase:
\be
\left.\frac{\delta\rho}{\bar{\rho}}\right\vert_{t = t_{\rm end-tran}} \sim  \frac{k^2}{H_0^2}\zeta \,,
\label{delrho1bis}
\ee
in agreement with the parametric dependence in~(\ref{eps3}). 

The growth in $\delta \rho/\bar{\rho}$ at first sight seems to contradict the attractor property established in Sec.~\ref{attract}. 
However, this is an artifact of $\bar{\rho}$ being accidentally small: large kinetic and potential energy contributions
nearly cancel on the background solution, resulting in a comparatively small total energy density. If we instead 
consider second-order contributions, such as $\delta\rho^{(2)}\sim \delta\dot{\phi}^2$, we obtain
\be
\frac{\delta\rho^{(2)}}{\delta\rho^{(1)}} \sim \frac{\delta\dot{\phi}^2}{\delta\dot{\phi}\;\dot{\phi}+ V_{,\phi}\delta\phi} \sim \frac{k^2t}{H_0}\,\zeta \,,
\label{delrho2}
\ee
which clearly becomes increasingly small in time, consistent with the attractor property of our solution. 
Note that evaluating~(\ref{delrho2}) at horizon crossing gives $\delta\rho^{(2)}/\delta\rho^{(1)} \sim k\zeta/|H_0|$. Hence
the perturbation expansion for $\delta \rho/\bar{\rho}$ breaks down for $k \;\gsim \; 10^5|H_0|$, consistent with
the $\dot{\eta}$ contribution to the three-point function --- see~(\ref{Adoteta}). It will be shown below that $10^5|H_0|$ also coincides with the onset
of strong coupling.

The authors of~\cite{mukhanovlinde} performed a similar analysis in {\it Newtonian} gauge and instead found 
\be
\left. \frac{\delta\rho}{\bar{\rho}}\right\vert_{\rm Newtonian} \sim c^2\Phi\,,
\label{delrhoLMV}
\ee
where $\Phi\sim 10^{-5}$ is the gravitational potential. If true, then for the values of $c$ of interest this would
invalidate perturbation theory on {\it all} scales. However, this is clearly an artifact of a poor gauge choice. As emphasized in~\cite{nicolis},
Newtonian gauge is ill-suited to study the evolution of perturbations in ekpyrotic cosmology, since $\delta\phi$ and $\Phi$ both
diverge as $1/t$. In the case of $\delta\rho/\bar{\rho}$, the relation between Newtonian and synchronous gauge is (using $a(t)\approx 1$)\footnote{We thank Alex Vikman and Guido D'Amico for discussions on this point.}
\bea
\nonumber
\left. \frac{\delta\rho}{\bar{\rho}}\right\vert_{\rm Newtonian} &=& \left. \frac{\delta\rho}{\bar{\rho}}\right\vert_{\rm sync} - \frac{\dot{\bar{\rho}}}{\bar{\rho}}\left[ \frac{\zeta}{H} -\frac{\epsilon}{\vec{\nabla}^2}\dot{\zeta} - \int^{t'}{\rm d}t''\frac{\dot{\zeta}}{H} \right] \\
\nonumber
&=&  \left. \frac{\delta\rho}{\bar{\rho}}\right\vert_{\rm sync} + 2\epsilon \zeta + \ldots  \\
& \xrightarrow[t=t_{\rm end-tran}]{} &  \left. \frac{\delta\rho}{\bar{\rho}}\right\vert_{\rm sync} + c^2\zeta + \ldots\,.
\eea
And since $\Phi = \zeta$ at the end of the transition phase~\cite{mukhanovlinde}, this is consistent with~(\ref{delrhoLMV}). The large contribution in~(\ref{delrhoLMV})
is therefore purely a consequence of a breakdown of Newtonian gauge. Similar conclusions apply to other quantities in the two gauges, such as $\delta\phi$.

\subsection{Quantum Corrections}
\label{quantumdiag}

Next we turn to quantum considerations and argue that the pure exponential case studied thus far is dominated
by quantum effects on small scales. Specifically, we will see that loop corrections to the two-point function
overwhelm the tree-level contribution, signaling strong coupling.

A quick estimate of the magnitude of loop corrections can be obtained by comparing the cubic and quadratic action for $\zeta$ at freeze out,
where quantum effects are most important~\cite{louissarah}. As shown in Sec.~\ref{horcrossvslw}, the dominant contribution at freeze out arises
from the $\dot{\eta}$ vertex:
\be
\left.\frac{S_3}{S_2}\right\vert_{k|t| = 1} \sim \frac{k}{|H_0|}\zeta\,.
\label{loopcond}
\ee
Thus the theory is strongly coupled for $k\;\gsim\; 10^5|H_0|$. 

On yet smaller scales, the stress tensor of quantum fluctuations dominates the background energy density, indicating a backreaction problem. This can be estimated by comparing
the quadratic action $S_2$ at horizon crossing with the background action $S_0\sim H_0^2M_{\rm Pl}^2$. This gives
\be
\left.\frac{S_2}{S_0}\right\vert_{k|t| = 1} \sim \left.\frac{\epsilon\,\dot{\zeta}^2}{H_0^2}\right\vert_{k|t| = 1} \sim \frac{k^4}{c^2H_0^4}\zeta^2 \sim \frac{k^4}{H_0^2M_{\rm Pl}^2} \,,
\label{backreactioncond}
\ee
where in the last step we have used $\zeta \sim c|H_0|/M_{\rm Pl}$ for the scale invariant modes generated during the transition phase. Hence this ratio is also
$\gg 1$ on sufficiently small scales. 

\subsection{Summary}

Let us briefly recap the issues uncovered in this Section. We have found that non-gaussianities are strongly scale dependent, resulting in a breakdown of
classical perturbation theory on small scales. The dominant contribution to $f_{\rm NL}$ comes the $\epsilon^3$ vertices in the cubic action, which, remarkably,
peaks at late times, well after the modes of interest have frozen out. As a result, the perturbative expansion parameter $f_{\rm NL}\zeta$
becomes larger than unity for
\be
k \;\gsim\; 10^{5/2}\,|H_0|\,.
\ee
On smaller scales, loop corrections eventually dominate the two-point function. Specifically, the theory is strongly coupled when modes with
\be
k \;\gsim\; 10^5\,|H_0|
\ee
are generated. Correspondingly, we have found that the (classical) perturbative expansion for $\delta\rho/\bar{\rho}$ breaks down on those scales.
On yet smaller scales, quantum backreaction effects overwhelm the background.

As mentioned earlier, these problems all result from maintaining the transition phase with large $c$ longer than necessary. We will see in the next Section
that strong coupling can be avoided by altering the potential such that $\zeta$ strongly tilts to the red on small scales. Suppressing the small-scale amplitude
in this way in turn allows perturbation theory to be valid on all scales, both classically and quantum mechanically.

\section{Weakly-Coupled Model}
\label{smoothmodel}

The aforementioned small-scale suppression of power can be achieved by generalizing~(\ref{potent}) to
\be
V(\phi) = V_0\left(1-e^{-c(\phi)\phi/M_{\rm Pl}}\right)\,,
\ee
where $c(\phi)$ decreases smoothly to $b\ll c$ after the transition phase has generated an acceptable range of scale invariant fluctuations. 
It is reasonable to expect that our results will depend on how rapidly $c(\phi)$ decreases and on its asymptotic value $b$, but should
otherwise be insensitive to the details of this process. Hence, instead of specifying $c(\phi)$ it is more convenient to choose
a suitable $\epsilon(t)$ that allows us to proceed analytically.

We require that $\epsilon(t) \approx 2/c^2H_0^2t^2$, corresponding to $c(\phi) = c$, from the onset of the transition phase
until some time $t_c$. Therefore, during the interval $t_{\rm beg-tran} < t < t_c$, the transition phase proceeds as before,
and scale invariant modes are generated with amplitude given by~(\ref{u}). This standard part of the evolution will be referred to as the scale invariant phase.
We assume that $\epsilon$ is continuous at $t= t_c$ and subsequently grows as a power-law:
\be
\epsilon(t) = -\frac{\dot{H}}{H^2} = \frac{6M_{\rm Pl}^2}{c^2V_0t_c^2}\left(\frac{t_c}{t}\right)^{2(1+\alpha)}\,,
\label{epssmooth}
\ee
where $\alpha > 0$. (The power-law form is convenient because the $\zeta$ mode function equation can be solved analytically in terms of Hankel functions.) 
We will refer to this phase as the $\alpha$ phase.
Meanwhile, since $H(t)\approx H_0$ during the transition phase, we can integrate the relation $\epsilon(t) = -\dot{H}/H^2$ to obtain
\be
H(t) \approx H_0 + \frac{2}{c^2t_c(1+2\alpha)}\left(\frac{t_c}{t}\right)^{1+2\alpha}\,.
%\qquad {\rm for}\;\; t_s < t < t_c \,.
\label{Halpha}
\ee
Paradoxically, $\epsilon(t)$ increases {\it faster} than in the pure exponentially case, which at first sight would seem to exacerbate the problems encountered earlier.
In fact, this is not so. A faster growth in $\epsilon$ can result in a {\it shorter} transition phase, which in turn implies a {\it smaller} value of $\epsilon$ at the onset
of the ekpyrotic scaling phase.

We assume that the $\alpha$ phase ends at a time $t_s$, at which time the universe enters an ekpyrotic scaling phase with $\epsilon = b^2/2$.  
Assuming continuity of $\epsilon$ at $t_s$,~(\ref{epssmooth}) implies
\be
b^2 =  \frac{12M_{\rm Pl}^2k_c^2}{c^2V_0}\left(\frac{k_s}{k_c}\right)^{2(1+\alpha)}\,,
\label{b^2reln}
\ee
where we have introduced the notation $k_c \equiv |t_c|^{-1}$ and $|k_s| \equiv |t_s|^{-1}$ for future convenience, corresponding
to the shortest-wavelength modes generated during the scale invariant and $\alpha$ phase, respectively. 

During the ekpyrotic phase ($t > t_s$), the Hubble parameter is given by
\be
H(t) = \frac{2}{b^2(t-t_{\rm crunch})}\qquad {\rm for}\;\;  t > t_s \,,
\label{Hekphase}
\ee
where $t_{\rm crunch}$ marks the time of the big crunch. (In reality, we of course envision that the ekpyrotic phase itself terminates before the big crunch and is followed
by a bounce to an expanding, radiation-dominated phase. In the New Ekpyrotic scenario~\cite{newek1}, for instance, a non-singular bounce is achieved through a ghost condensate~\cite{nicolisbounce}.
See~\cite{jeanlucsusy} for a recent supersymmetric extension of this theory.) Matching~(\ref{Halpha}) and~(\ref{Hekphase}) at $t_s$ gives
\be
t_{\rm crunch} = t_s - \frac{2}{b^2H_0}\,.
\label{tcrunch}
\ee

Before turning our attention to perturbations, we note in passing that the modified evolution described above can circumvent one of the criticisms raised in~\cite{mukhanovlinde}, namely that
$\dot{H},\ddot{H},\ldots$ all eventually become super-Planckian in the pure exponential case. Indeed, at the end of the end of the ekpyrotic scaling phase, we have $\dot{H}_{\rm ek-end} = c^2H^2_{\rm ek-end}/2$, which is $\gg M_{\rm Pl}^2$ for the values of $c$ considered here. Higher derivatives of $H$ are even more singular:
\be
\left\vert \frac{{\rm d}^nH}{{\rm d}t^n}\right\vert^{\frac{1}{n+1}} \sim c^{\frac{2n}{n+1}}|H_{\rm ek-end}|
\xrightarrow[n\rightarrow\infty]{}  c^2 |H_{\rm ek-end}| \gg M_{\rm Pl}\,.
\label{Hder}
\ee
With the modified evolution, however, the universe eventually matches on to an ekpyrotic scaling phase with a much smaller $\epsilon$.
All time-derivatives of $H$ will remain sub-Planckian provided that $b^2 |H_{\rm ek-end}| < M_{\rm Pl}$. But $b$ must also be
$\gsim\; 1$, since an ekpyrotic phase, by definition, corresponds to an equation of state parameter larger than unity. In other words, the allowed range is
\be
1< b^2 < \frac{M_{\rm Pl}}{|H_{\rm ek-end}|}\,,
\label{b^2bound}
\ee
which can be satisfied for a wide range of parameters.

\subsection{Mode Functions}
\label{pertanalysis}

Next we solve for the curvature perturbation, tracking its evolution throughout the $\alpha$ phase and subsequent ekpyrotic scaling phase.

\noindent {\bf $\alpha$ phase}: During the intermediate phase in which $\epsilon$ evolves as~(\ref{epssmooth}), the scale factor remains nearly constant, and hence 
$z\equiv a\sqrt{2\epsilon} \sim 1/|t|^{1+\alpha}$. The evolution equation~(\ref{vpert}) therefore reduces to
\be
\ddot{v}_k + \left(k^2 - \frac{(1+\alpha)(2+\alpha)}{t^2}\right)v_k = 0 \qquad {\rm for}\;\; t_c < t < t_s \,.
\label{vpertalp}
\ee
Let us first discuss modes that freeze out during the scale invariant phase ($t < t_c$), {\it i.e.} those with $k < k_c$. These modes are already frozen out
by the onset of the modified transition phase, hence~(\ref{zetalw}) applies just before $t=t_c$:
\be
\zeta_{k<k_c}(t < t_c) =  \frac{i c |H_0|}{2\sqrt{2}M_{\rm Pl}k^{3/2}}  \left(1 + \frac{1}{2}k^2t^2 + \frac{i}{3}k^3t^3 + \ldots \right)\,.
\label{zetalwt<tc}
\ee
Here we have used $|t| > |t_c| \gg |t_{\rm end-tran}|$.
By comparing~(\ref{vpertexact}) and~(\ref{vpertalp}), we note that for $\alpha\sim {\cal O}(1)$, the freeze-out radius $|H_{\rm freeze}|^{-1} = \sqrt{z/\ddot{z}}$ only changes by a factor of order unity
at $t_c$; hence modes with $k < k_c$ do not re-enter the freeze-out horizon. Thus, right after $t=t_c$, we can solve~(\ref{vpertalp}) in the long
wavelength limit
\be
\zeta_{k<k_c}(t > t_c) = A_k \left[ 1+ \frac{k^2t^2}{2(1+2\alpha)} + \ldots \right] + B_k \left[(-kt)^{3+2\alpha} + \ldots\right]\,.
\label{zetalwt>tc}
\ee
Matching $\zeta$ and $\dot{\zeta}$ at $t = t_c$ allows us to fix $A_k$ and $B_k$. The relevant terms are
\be
k^{3/2}\zeta_{k<k_c}  \simeq  \frac{i c |H_0|}{2\sqrt{2}M_{\rm Pl}} \left\{ 1+ \frac{k^2t^2}{2(1+2\alpha)} + \frac{i}{3+2\alpha} \left(\frac{k_c}{k}\right)^{2\alpha} (-kt)^{3+2\alpha}\right\}\,.
\label{lwmodexp}
\ee
These modes therefore remain scale invariant throughout the $\alpha$ phase.

Modes with $k> k_c$, on the other hand, are still in their adiabatic vacuum at the onset of the $\alpha$ phase. With this vacuum choice, the mode function solution is
\be
v_{k>k_c} = \frac{\sqrt{-\pi t}}{2M_{\rm Pl}} H_{\alpha+3/2}^{(1)}(-kt)\,.
\label{vinterm}
\ee
Using the asymptotic expansion of the Hankel function, the long-wavelength curvature perturbation $\zeta_k = v_k/z$ on these scales is
\be
k^{3/2}\zeta_{k>k_c} = \frac{i c |H_0|}{2\sqrt{2}M_{\rm Pl}} \left(\frac{k_c}{k}\right)^{\alpha}\frac{2^{1+\alpha}\Gamma(\alpha + 3/2)}{\sqrt{\pi}} \Bigg\{ 1+ \frac{k^2t^2}{2(1+2\alpha)} + \frac{i\pi  (-kt)^{3+2\alpha}}{2^{2(1+\alpha)}(3+2\alpha)\Gamma^2(\alpha+3/2)}\Bigg\}\,.
\label{swmodexp}
\ee
Hence the spectrum has a strong red tilt for $\alpha \sim {\cal O}(1)$, as desired. 

\noindent {\bf Ekpyrotic scaling phase}: During the ekpyrotic scaling phase ($t > t_s$), the equation of state is nearly constant and large, $\epsilon = b^2/2\gg 1$, and hence
the scale factor slowly contracts as power-law, $a(t) \sim (-t)^{2/b^2}$. The evolution equation~(\ref{vpert}) in this case reduces to
\be
\ddot{v}_k + \left(k^2 - \frac{2}{b^2(t-t_{\rm crunch})^2}\right)v_k = 0 \qquad {\rm for}\;\; t > t_s \,.
\label{vpertek}
\ee
Unlike the scale invariant to $\alpha$ phase transition, the $\alpha$ to ekpyrotic scaling transition typically implies a substantial change in the
freeze-out horizon. From~(\ref{vpertek}), the freeze-out horizon at the onset of the ekpyrotic scaling phase is
\be
H_\zeta^{-1} \big\vert_{t = t_s^+} = b\left\vert t_s - t_{\rm crunch}\right\vert = \frac{1}{b|H_0|}\,,
\ee
where we have used~(\ref{tcrunch}). On the other hand, for $\alpha\sim {\cal O}(1)$, we can read off from~(\ref{vpertalp}) that
\be
H_\zeta^{-1} \big\vert_{t = t_s^-}  \simeq |t_s|\,.
\ee
For our parameter choices, we will see that $H_\zeta^{-1} \big\vert_{t = t_s^-}/H_\zeta^{-1} \big\vert_{t = t_s^+} = bH_0t_s \ll 1$, hence some of the modes generated
during the scale invariant and $\alpha$ phases re-enter the freeze-out at $t=t_s$. We must therefore carefully keep track of their evolution.

The general solution to~(\ref{vpertek}) is
\bea
\nonumber
v_k(t > t_s) &=& \sqrt{k(-t)}\left[A_kJ_{\frac{1}{2}\sqrt{\frac{b^2-8}{b^2}}}(k(t_{\rm crunch}-t)) + B_k Y_{\frac{1}{2}\sqrt{\frac{b^2-8}{c^2}}}(k(t_{\rm crunch}-t))\right] \\
\nonumber
&\approx & \sqrt{k(-t)}\bigg[A_kJ_{1/2}(k(t_{\rm crunch}-t)) + B_k Y_{1/2}(k(t_{\rm crunch}-t))\bigg] \\
&=& \sqrt{\frac{2}{\pi}}\bigg[A_k \sin(k(t_{\rm crunch}-t)) - B_k\cos(k(t_{\rm crunch}-t))\bigg]\,,
\label{veksol}
\eea
where the second step follows because $b\gg 1$. And since $z = a\sqrt{2\epsilon}\approx b$ is constant in this phase, the curvature perturbation is simply given by
\be
\zeta_k (t > t_s) = b^{-1}\sqrt{\frac{2}{\pi}}\bigg[A_k \sin(k(t_{\rm crunch}-t)) - B_k\cos(k(t_{\rm crunch}-t))\bigg]\,.
\ee
Matching this to~(\ref{lwmodexp}) and~(\ref{swmodexp}), respectively, and using the fact that $k < k_s$ for the modes of interest, we obtain at late times ($k|t_{\rm crunch} - t| \ll 1$):
\bea
\nonumber
k^{3/2} \zeta_k &\simeq & \frac{i c |H_0|}{4\sqrt{2}M_{\rm Pl}} \cos\left(\frac{2k}{b^2|H_0|}\right) \;\;\;\;\;\;\;\;\;\;\;\;\;\;\;\;\;\;\;\;\;\;\;\;\;\;\;\;\;\;\;\;\;\;\;\;\;\;\;\;\;\;\;\, {\rm for}\;\;\; |H_0| < k < k_c\;; \\
k^{3/2} \zeta_k &\simeq &  \frac{i c |H_0|}{4\sqrt{2}M_{\rm Pl}} \left(\frac{k_c}{k}\right)^{\alpha}\frac{2^{1+\alpha}\Gamma(\alpha + 3/2)}{\sqrt{\pi}}  \cos\left(\frac{2k}{b^2|H_0|}\right) \qquad {\rm for} \;\;\; k_c < k < k_s\,.
\label{zetampfinalfinal}
\eea
Therefore, aside from acquiring an oscillatory factor, $\zeta$ maintains its original amplitude throughout the ekpyrotic scaling phase. The cosine factor results in oscillations in the power
spectrum. For this effect to be negligible on the largest scales probed by observations, we demand that the cosine be approximately constant over the entire scale invariant range. This will be the case if
\be
\frac{k_c}{b^2|H_0|} < 1\,.
\label{kcb^2bound}
\ee
On small scales, $\zeta$ has a strong red tilt, and assumes a minimum amplitude for the shortest-wavelength mode ($k = k_s$) generated during the $\alpha$ phase:
\be
k^{3/2}|\zeta_k|_{\rm min} \sim \frac{c|H_0|}{M_{\rm Pl}} \left(\frac{k_c}{k_s}\right)^\alpha\,.
\label{zetamin}
\ee
Finally, on yet even smaller scales, modes with $k > k_s$ freeze out during the ekpyrotic scaling phase, and as usual have a strong blue tilt.
Imposing the adiabatic vacuum choice in~(\ref{veksol}) fixes $A_k,B_k\sim 1/\sqrt{2k}M_{\rm Pl}$, hence
\be
k^{3/2}\zeta_k \sim \frac{k}{bM_{\rm Pl}} \qquad {\rm for}\;\;\; k > k_s\,.
\ee
This growth is cut off once the ekpyrotic phase terminates, which occurs well before the amplitude reaches unity.

\subsection{Avoiding Strong Coupling}
\label{scavoid}

The strong red tilt on intermediate scales generated during the $\alpha$ phase can cure the strong coupling problem encountered in Sec.~\ref{quantumdiag}.
The dominant contribution to $S_3/S_2$ at horizon crossing is given as before by~(\ref{loopcond}):
\be
\left.\frac{S_3}{S_2}\right\vert_{k|t| = 1} \sim \frac{k}{|H_0|}\zeta\,.
\label{loopcondbis}
\ee
On the largest scales ($k<k_c$), $\zeta$ is scale invariant and $\sim 10^{-5}$ as before. On smaller scales ($k>k_c$), however, we have
\be
\left.\frac{S_3}{S_2}\right\vert_{k|t| = 1} \sim \frac{k}{|H_0|}\zeta \sim k^{1-\alpha}\,.
\label{loopinterm}
\ee
For $\alpha > 1$, in particular, the theory becomes increasingly {\it weakly} coupled on small scales. Hence, provided that the range of scale
invariant modes satisfies
\be
k_c \;\lsim \; 10^5\,|H_0|\,,
\label{kcscbound}
\ee
then for $\alpha > 1$ quantum corrections are under control on all scales. This is our main constraint on the allowed range of scale invariant modes.

The $\alpha$ phase also allows us to circumvent the quantum backreaction problem of the pure exponential case. As in~(\ref{backreactioncond}), the backreaction
is largest on small scales, hence we focus on the modes generated during the $\alpha$ phase:
\be
\left.\frac{S_2}{S_0}\right\vert_{k|t| = 1} \sim \left.\frac{\epsilon\,\dot{\zeta}^2}{H_0^2}\right\vert_{k|t| = 1} \sim \frac{k^2k_c^2}{c^2H_0^4}\left(\frac{k}{k_c}\right)^{2(1+\alpha)}\zeta^2
\sim \frac{k^4}{H_0^2M_{\rm Pl}^2}\,,
\ee
where in the last step we have substituted~(\ref{zetampfinalfinal}). Although the parametric dependence is identical to~(\ref{backreactioncond}), the
upshot of the $\alpha$ phase is that it limits the range of modes generated. Backreaction peaks at $k=k_s$ and is under control provided that
\be
k_s^2 < |H_0|M_{\rm Pl} \,.
\label{ksscbound}
\ee

\subsection{Non-Gaussianities}
\label{NGimproved}

The modified evolution for $\epsilon(t)$ should have a dramatic impact on the three-point function. Indeed, recall that the dominant $\epsilon^3$ contribution
peaked at late times, which is precisely what has been altered with the $\alpha$ phase.

To see how non-linearities can be tamed, let us focus on the scale invariant modes ($|H_0| < k < k_c$). During the $\alpha$ phase, their evolution is described by~(\ref{lwmodexp}). 
As in Sec.~\ref{3ptcalc}, the dominant contribution to~(\ref{eps33pt}) comes from the imaginary part of the integrand. Substituting~(\ref{lwmodexp}), we obtain
\bea
\nonumber
& & \left(\frac{k_1^2}{k_2^2}\zeta_{k_1}^*\,\dot{\zeta}_{k_2}^* + 2\dot{\zeta}_{k_1}^*\zeta_{k_2}^* \right)\frac{\vec{k}_2\cdot \vec{k}_3}{k_3^2}\dot{\zeta}_{k_3}^* + {\rm perm.} = 
-\left(\prod_j  \frac{(-i) c |H_0|}{2\sqrt{2}M_{\rm Pl}k_j^{3/2}} \right) \\
& & \;\;\;\;\;\;\;\;\;\;\; \times \left\{\frac{2i}{1+2\alpha} K^2 \left(\sum_ik_i^3 - \sum_{i\neq j}k_ik_j^2+2k_1k_2k_3\right) |t_c|^{-2\alpha} (-t)^{3+2\alpha} + {\rm real\;part} \large\right\}\,.
\label{imeps3mod}
\eea
While this is suppressed for small $t$, the integral is once again overwhelmed by the growth in the vertex: $\epsilon^3 \sim 1/t^{6(1+\alpha)}$. The three-point function is
\bea
\nonumber
\langle \zeta(\textbf{k}_1)\zeta(\textbf{k}_2)\zeta(\textbf{k}_3)\rangle_{\epsilon^3} &=&  -\frac{(2\pi)^3\delta^3\left(\sum_i\kk_i\right)}{128\prod_j k_j^3}K^2\left(\sum_ik_i^3 - \sum_{i\neq j}k_ik_j^2+2k_1k_2k_3\right)   \\
\nonumber
&\times &   \frac{t_c^{4\alpha}}{1+2\alpha} \int^{t_s} \frac{{\rm d}t}{t^{3+4\alpha}} + {\rm c.c.} \\
&=& \frac{(2\pi)^3\delta^3\left(\sum_i\kk_i\right)}{128\prod_j k_j^3} \frac{K^2t_c^{2\alpha}}{(1+2\alpha)^2t_s^{2(1+2\alpha)}} \left(\sum_ik_i^3 - \sum_{i\neq j}k_ik_j^2+2k_1k_2k_3\right).
\eea 
Rewriting this in terms of $k_c = 1/|t_c|$ and $k_s = 1/|t_s|$, we find the amplitude
\be
{\cal A}_{\epsilon^3} = \frac{9M_{\rm Pl}^4}{8c^4V_0^2}\frac{k_c^2K^2}{(1+2\alpha)^2}  \left(\sum_ik_i^3 - \sum_{i\neq j}k_ik_j^2+2k_1k_2k_3\right)\left(\frac{k_s}{k_c}\right)^{2(1+2\alpha)}\,,
\label{eps3mod}
\ee
with corresponding $f_{\rm NL}^{\rm equil.}$ parameter:
\be
f_{\rm NL}^{\;\epsilon^3} \sim \frac{k_c^2k^2M_{\rm Pl}^4}{c^4V_0^2} \left(\frac{k_s}{k_c}\right)^{2(1+2\alpha)}\,.
\label{fNLeps3mod}
\ee
Note that, remarkably, if the $\alpha$ phase is maintained long enough to the point where the approximation $H\approx H_0$ breaks down, that is, if $t_s$ is chosen such that $|H_0|\sim 2c^{-2}k_c(1+2\alpha) (t_c/t_s)^{1+2\alpha}$, then~(\ref{fNLeps3mod}) exactly matches our earlier result for the pure exponential case: $f_{\rm NL}^{\;\epsilon^3} \sim k^2/H_0^2$.
{\it Therefore, by terminating the $\alpha$ phase at an earlier time, we can suppress $f_{\rm NL}^{\;\epsilon^3}$, as desired.}
Specifically, we demand that $f_{\rm NL}\zeta < 1$. For the scale invariant modes, $|H_0| < k < k_c$, the amplitude is $\zeta\sim 10^{-5}$, and the condition is most stringent at $k=k_c$:
\bea
\nonumber
\left(f_{\rm NL}\zeta\right)_{\epsilon^3}\big\vert_{k=k_c} &\simeq& 10^{-5} \frac{k_c^4M_{\rm Pl}^4}{c^4V_0^2} \left(\frac{k_s}{k_c}\right)^{2(1+2\alpha)} \\
&\simeq & 10^{15} \left(\frac{k_c}{H_0}\right)^4\left(\frac{H_0}{M_{\rm Pl}}\right)^4  \left(\frac{k_s}{k_c}\right)^{2(1+2\alpha)}  < 1\,,
\label{eps3condlargescale}
\eea
where in the last step we have used~(\ref{ampcons}). 

It turns out that a similar calculation for the $k_s < k < k_c$ modes, obtained by substituting the mode function~(\ref{swmodexp}) into the three-point amplitude, yields an identical expression for $f_{\rm NL}^{\epsilon^3}$. Since $k^{3/2}\zeta \sim k^{-\alpha}$ in this case --- see~(\ref{zetampfinalfinal}) ---, it follows that $(f_{\rm NL}\zeta)_{\epsilon^3}\sim k^{2-\alpha}$ peaks at $k=k_c$ if $\alpha > 2$, and hence is automatically $< 1$ when~(\ref{eps3condlargescale}) is satisfied. For simplicity, we will therefore impose
\be
\alpha > 2\,.
\label{alp>2}
\ee

The remaining contributions to the three-point function can be estimated by the horizon-crossing approximation, as in Sec.~\ref{horcrossvslw}. Since $\eta \sim 1/t$ during both the scale invariant and $\alpha$ phases,
we have $f_{\rm NL}^{\dot{\eta}} \sim k/|H_0|$ on all scales. As in the strong coupling discussion of Sec.~\ref{scavoid}, $f_{\rm NL}\zeta$ peaks at $k= k_c$ for $\alpha >1$, and is $<1$ provided~(\ref{kcscbound}) is satisfied. As before, the $\epsilon^2$ contributions are subdominant and can be neglected.

\subsection{Summary of Constraints}
\label{improvedcons}

To be phenomenologically viable, the generalized model described above must satisfy the following list of constraints:

\noindent {\bf 1. Correct large scale amplitude}: The amplitude of the long-wavelength modes ($k < k_c$) should
match observations of the large-scale power spectrum. From~(\ref{ampcons}) we have
\be
\frac{c|H_0|}{M_{\rm Pl}} = 10^{-5}\,.
\label{amp}
\ee

\noindent {\bf 2. Scale invariant modes must match the observable range}: The modes generated during the scale invariant phase are on scales smaller than $|H_0|^{-1}$, hence the
comoving scale  $|H_0|^{-1}$ must encompass the entire observable universe. This leads to the upper bound on $|H_0|$ given by~(\ref{V0cons}), where recall that $H_{\rm ek-end}$,
the Hubble parameter at the end of the ekpyrotic scaling phase, is assumed comparable in magnitude to the Hubble parameter at the onset of the expanding, radiation-dominated phase:
$|H_{\rm ek-end}| \sim T_{\rm reheat}^2/M_{\rm Pl}$. For simplicity, we will assume that the bound~(\ref{V0cons}) is saturated, thereby fixing $|H_0|$ in terms of the reheating scale:
\be
|H_0| = 10^{-30}\;T_{\rm reheat}\,.
\label{V0consbis}
\ee

\noindent {\bf 3. Avoiding strong coupling}: As discussed in Sec.~\ref{scavoid}, loop corrections are small if the range of scale invariant modes is restricted
to $k_c \;\lsim \; 10^5\,|H_0|$ --- see~(\ref{kcscbound}). (Another necessary condition is $\alpha > 1$, but this now follows from~(\ref{alp>2}).) Since a factor of $10^5$ is just enough to be
consistent with microwave background and large scale structure observations, we fix $k_c$ to nearly saturate this bound
\be
k_c \simeq 10^5\,|H_0|\,.
\label{kcfixed}
\ee

Meanwhile, quantum backreaction is under control provided that $k_s^2<|H_0|M_{\rm Pl}$ --- see~(\ref{ksscbound}). Using~(\ref{V0consbis}) and~(\ref{kcfixed}), we can rewrite this as
\be
\frac{k_c}{k_s}  > 10^{-10}\sqrt{\frac{T_{\rm reheat}}{M_{\rm Pl}}}\,.
\label{kcks1}
\ee

\noindent {\bf 4. Scale invariance and sub-Planckian curvatures}: The equation of state during the ekpyrotic phase, $\epsilon = b^2/2$, is
constrained by two requirements. To avoid reaching super-Planckian curvatures by the end of the ekpyrotic phase, $b^2$ is bounded from above through~(\ref{b^2bound}).
Given our assumption that $|H_{\rm ek-end}|$ sets the reheating scale, the upper bound implies $b^2 < M_{\rm Pl}^2/T_{\rm reheat}^2$. Substituting~(\ref{b^2reln})
and using the relations~(\ref{amp}) and~(\ref{V0consbis}), we can rewrite this inequality as
\be
\frac{k_c}{k_s} > \left[ 10^{-20}\left(\frac{T_{\rm reheat}}{M_{\rm Pl}}\right)^2\right]^{\frac{1}{1+\alpha}}\,.
\label{kcks2}
\ee
The equation of state is also bounded from below by demanding that the scale invariant spectrum is not appreciably distorted by the ekpyrotic scaling phase.
From~(\ref{kcb^2bound}) and~(\ref{kcfixed}), this requires $b^2 > 10^5$. Once again substituting~(\ref{b^2reln}) etc., we obtain
\be
\frac{k_c}{k_s} < \left[ 10^{-45}\left(\frac{T_{\rm reheat}}{M_{\rm Pl}}\right)^2\right]^{\frac{1}{2(1+\alpha)}}\,.
\label{kcks3}
\ee

\noindent {\bf 5. Validity of classical perturbation theory}: We have seen in Sec.~\ref{NGimproved} that, provided $\alpha > 2$, the tightest constraint from non-gaussianities
comes from the $\epsilon^3$ contribution at $k=k_c$. Using~(\ref{V0consbis}) and~(\ref{kcfixed}), the inequality~(\ref{eps3condlargescale}) can be rewritten as a constraint on $k_c/k_s$:
\be
\frac{k_c}{k_s} > \left[ 10^{-85}\left(\frac{T_{\rm reheat}}{M_{\rm Pl}}\right)^4\right]^{\frac{1}{2(1+2\alpha)}}\,.
\label{kcks4}
\ee
This is generally a more stringent lower bound than either~(\ref{kcks1}) or~(\ref{kcks2}).

\subsection{Working Examples}
\label{working}

To summarize, once we choose a reheating scale $T_{\rm reheat}$, the scale of the potential $|H_0|$ (or equivalently, $V_0$) is fixed by~(\ref{V0consbis}). In turn, the exponent $c$ characterizing the
scale invariant phase is fixed by the large scale normalization~(\ref{amp}), while the comoving scale $k_c$ marking the onset of the $\alpha$ phase is determined by~(\ref{kcfixed}). Only two
parameters remain to be specified: $\alpha$ and $k_c/k_s$. For a given $\alpha > 2$, we will check that $k_c/k_s$ satisfies the inequalities~(\ref{kcks1})$-$(\ref{kcks4}). 

\noindent {\bf High-Scale Example:} Consider GUT-scale reheating, $T_{\rm reheat} = 10^{-3}\;M_{\rm Pl} = 10^{15}\;{\rm GeV}$. From~(\ref{amp}) and~(\ref{V0consbis}) this
fixes $V_0^{1/4} = 10\;{\rm GeV}$ and $c = 10^{28}$. Choosing $\alpha = 5$, the upper bound~(\ref{kcks3}) reduces to $k_c/k_s \;\lsim\;  5\times 10^{-5}$. Meanwhile, among the lower bounds,~(\ref{kcks4}) is the most stringent: $k_c/k_s \;\gsim\; 4\times 10^{-5}$. Dropping factors of order unity, we therefore impose
\be
\frac{k_c}{k_s} = 10^{-5}\,.
\ee
Hence $\simeq 12$ e-folds of modes are generated during the $\alpha$ phase in this case.

\noindent {\bf Intermediate-Scale Example:} Consider reheating at an intermediate scale, $T_{\rm reheat} = 10^{-9}\;M_{\rm Pl} = 10^{9}\;{\rm GeV}$, corresponding to
$V_0^{1/4} = 10^{-2}\;{\rm GeV}$ and $c = 10^{34}$. Choosing $\alpha = 3$, the upper bound~(\ref{kcks3}) reduces to $k_c/k_s \;\lsim\; 10^{-8}$, while the
tightest lower bound is again given by~(\ref{kcks4}): $k_c/k_s \;\gsim\;  8\times 10^{-9}$. Hence in this case we impose
\be
\frac{k_c}{k_s} = 10^{-8}\,,
\ee
corresponding to $\simeq 18$ e-folds of $\alpha$-phase modes.

\noindent {\bf Low-Scale Example:} Consider electroweak-scale reheating, $T_{\rm reheat} = 10^{-15}\;M_{\rm Pl} = 10^{3}\;{\rm GeV}$, corresponding to 
$V_0^{1/4} = 10^{-5}\;{\rm GeV}$ and $c = 10^{40}$. With $\alpha = 3$, we find that the inequalities on $k_c/k_s$ are satisfied for
\be
\frac{k_c}{k_s} = 10^{-10}\,,
\ee
which amounts to $\simeq 23$ e-folds of $\alpha$-phase modes.

\section{Conclusions}
\label{conclu}

In this paper, we have explored the adiabatic ekpyrotic mechanism proposed recently to generate a scale invariant
spectrum within an attractor background. At the level of the power spectrum, the adiabatic mechanism is  dual to inflation --- the equation governing
$\zeta$ and its growing mode solution are identical to inflationary cosmology. 

As we have seen, however, the duality is broken by the three-point correlation function. Unlike the nearly gaussian spectrum of inflation, the
rapidly-varying equation of state $\epsilon\sim 1/\tau^2$ characteristic of the adiabatic ekpyrotic phase 
results in large non-gaussianities on small scales. For the simplest exponential potential~(\ref{potent}), the most dangerous contribution
comes from ${\cal O}(\epsilon^3)$ terms in the cubic action, which are subdominant in the inflationary case. 
The rapid growth in these vertices gives three-point contributions that peak at late times when $\epsilon \gg 1$.
At the same time, loop corrections dominate the tree-level computation on small scales.

This strong coupling and perturbative breakdown both trace back to the fact that the transition phase with large $c$ is
maintained longer than necessary. As we have seen, these pathologies can be avoided by considering more general potentials where the exponent decreases
smoothly from $c$ to a much smaller value $b\ll c$ once a suitable range of scale invariant modes has been generated. This suppresses
power on small scales, and thereby restores the validity of perturbation theory on all scales. We have shown that the resulting range of
nearly scale invariant and gaussian modes can span at most a factor of $10^5$ in $k$ space, which is enough to account for microwave
background and large scale structure observations. 

We are currently generalizing the scenario to the case of time-dependent sound speed $c_s(\tau)$, as expected in non-canonical scalar field theories,
with the hope that this can alleviate the issue of non-linearities. As shown in~\cite{piazza}, there is much more freedom in generating scale invariant perturbations in this case:
for any background $a(\tau)$ there exists in principle a suitable $c_s(\tau)$ such that $\zeta$ acquires a scale invariant spectrum. It will be interesting to see if $c_s(\tau)$ can also
tame the growth in the three-point function.

{\it Acknowledgments:} We thank Daniel Baumann, Tom~Banks, Guido~D'Amico, Austin~Joyce, Jean-Luc~Lehners, Godfrey~Miller, Alberto~Nicolis, Burt~Ovrut, Andrew~Tolley, Neil~Turok, Alex~Vikman, Daniel Wesley and Matias Zaldarriaga for helpful discussions. This work was supported in part by the US Department of Energy grant DE-FG02-91ER40671 (PJS) and by funds from the University of Pennsylvania and the Alfred P. Sloan Foundation (JK).

\section{Appendix: Calculation of three-point Function}
\label{apA}

In this Appendix we complete the calculation of Sec.~\ref{NG} and compute all remaining contributions to the three-point function
for the lifted exponential potential~(\ref{potent}). As in Sec.~\ref{NG}, we assume $M_{\rm Pl} = 1$ throughout. The contributions listed below refer to the cubic action~(\ref{action3}).

\begin{itemize}

\item {\bf The $\zeta \dot\zeta^2$ contribution}: This interaction Hamiltonian in this case is $H_{\rm int}  = -\int {\rm d}^3x \epsilon^2\,\zeta\dot{\zeta}^2$. Following similar steps as in Sec.~\ref{3ptcalc}, we obtain
\bea
\nonumber
& & \langle \zeta(\textbf{k}_1)\zeta(\textbf{k}_2)\zeta(\textbf{k}_3)\rangle_{\zeta \dot\zeta^2}\, =\, i (2 \pi)^3 \delta^3(\kk_1+\kk_2+\kk_3) \frac{2}{\prod_j k_j^3} \left(\frac{c^2H_0^2}{8}\right)^3\prod_j(1+ik_jt_{\rm end-tran})\\
\nonumber
& & \;\;\;\;\;\;\times \sum_{i<j} \int_{-\infty+i\varepsilon}^{0} {\rm d}t\, \epsilon^2(t)\,(t+t_{\rm end-tran})^2\left[k_i^2k_j^2 e^{iKt}-ik_1k_2k_3(t+t_{\rm end-tran})k_ik_je^{iKt} \right] + {\rm c.c.} \\
&& \;\;\;\;\;\;\;\approx -(2 \pi)^3 \delta^3(\kk_1+\kk_2+\kk_3)  \frac{c^2H_0^2}{32\cdot \prod_j k_j^3}  \sum_{i\neq j}k_i^2k_j^3\,\log(K|t_{\rm end-tran}|) \,,
\label{zetadot1}
\eea
where in the last step we have assumed $K|t_{\rm end-tran}|\ll 1$, appropriate for the modes of interest, and where we have used the identity
\be
K\sum_{i<j} k_i^2k_j^2 - k_1k_2k_3\sum_{i<j}k_ik_j = \sum_{i\neq j}k_i^2k_j^3\,.
\ee
The corresponding amplitude is 
\be
{\cal A}_{\zeta \dot\zeta^2} \approx -\frac{1}{2c^2H_0^2}\sum_{i\neq j}k_i^2k_j^3\,\log(K|t_{\rm end-tran}|)\,.
\label{Azetadotzeta}
\ee
This contribution is suppressed by $1/c^2$ relative to the dominant, $\epsilon^3$ amplitude, given by~(\ref{eps3}).

\item {\bf The $\zeta (\vec{\nabla}\zeta)^2$ contribution}: The interaction Hamiltonian in this case is $H_{\rm int} = -\int {\rm d}^3x\,\epsilon^2\zeta (\vec{\nabla}\zeta)^2$, with corresponding three-point function
\bea
\nonumber
& &\langle \zeta(\textbf{k}_1)\zeta(\textbf{k}_2)\zeta(\textbf{k}_3)\rangle_{\zeta (\vec{\nabla}\zeta)^2}  =  -(2 \pi)^3 \delta^3(\kk_1+\kk_2+\kk_3) \frac{c^2H_0^2\sum_i k_i^2}{64\cdot \prod_j k_j^3} \; {\rm Im} \left\{ \prod_j(1+ik_jt_{\rm end-tran}) \right. \\
\nonumber
& &  \left. \times  \int_{-\infty+i\varepsilon}^{0}\frac{{\rm d}t\;e^{iKt}}{(t+t_{\rm end-tran})^4}\left( 1-iKt   - \sum_{i<j}k_ik_j(t+t_{\rm end-tran})^2 +ik_1k_2k_3(t+t_{\rm end-tran})^3 \right)\right\}\,, \\
\eea
where we have used the identity
\be
\kk_1\cdot\kk_2 + \kk_2\cdot\kk_3 + \kk_1\cdot\kk_3 = -\frac{1}{2}\sum_i k_i^2\,.
\ee
Performing the integrals in the $K|t_{\rm end-tran}|\ll 1$ limit, we obtain the amplitude 
\be
{\cal A}_{\zeta (\vec{\nabla}\zeta)^2} \approx -\frac{1}{12c^2H_0^2} \sum_i k_i^2\sum_jk_j^3 \log(K|t_{\rm end-tran}|) \,.
\label{Avecnabla}
\ee
This is also suppressed by $1/c^2$ relative to the $\epsilon^3$ contribution.

\item{\bf The $\dot{\zeta}\vec{\nabla} \zeta\vec{\nabla} \chi$ contribution}: From the definition of $\chi$ in~(\ref{chidef}), we find $H_{\rm int} = 2\epsilon^2 \int {\rm d}^3x\,\dot{\zeta}\vec{\nabla}\zeta\frac{\vec{\nabla}}{\nabla^2}\dot{\zeta}$. The three-point contribution is therefore given by
\bea
\nonumber
& & \langle \zeta(\textbf{k}_1)\zeta(\textbf{k}_2)\zeta(\textbf{k}_3)\rangle_{\dot{\zeta}\vec{\nabla} \zeta \vec{\nabla} \chi} = (2 \pi)^3 \delta^3(\kk_1+\kk_2+\kk_3) \frac{c^2H_0^2}{32\cdot \prod_jk_j^3}\; {\rm Im} \left\{ \prod_j(1+ik_jt_{\rm end-tran}) \right. \\
& & \left. \;\;\;\;\;\;\;\;\; \times \int_{-\infty+i\varepsilon}^{0} \frac{{\rm d}t}{(t+t_{\rm end-tran})^2} \left(\kk_2\cdot\kk_3k_1^2[1-ik_2(t+t_{\rm end-tran})]e^{iKt}+{\rm perms.}\right)\right\} \,.
\eea
Performing the integrals and using the identities 
\bea
\nonumber
\frac{k_1^2\kk_2\cdot\kk_3}{K} +{\rm perms.} &=& \sum_ik_i^3-\sum_{i\neq j}k_ik_j^2+2k_1k_2k_3 \\
-\frac{k_1^2k_2\kk_2\cdot\kk_3}{K^2} +{\rm perms.} &=& -\frac{1}{2}\sum_{i\neq j}k_ik_j^2+2k_1k_2k_3 + \frac{2}{K^2}\sum_{i\neq j}k_i^2k_j^3\,,
\eea
we find the following leading piece for $K|t_{\rm end-tran}|\ll 1$
\be
{\cal A}_{\dot{\zeta}\vec{\nabla} \zeta \vec{\nabla} \chi} = \frac{K^2}{2c^2H_0^2}\left(\sum_ik_i^3-\frac{3}{2}\sum_{i\neq j}k_ik_j^2+\frac{2}{K^2}\sum_{i\neq j}k_i^2k_j^3 + 4k_1k_2k_3\right)\log(K|t_{\rm end-tran}|)\,.
\label{Amessyterm}
\ee
Again there is a $1/c^2$ suppression compared to the dominant contribution.

\item{\bf Field Redefinition}: The three-point function also receives contributions from the field definition~(\ref{redef}):
\bea
\zeta &\rightarrow & \zeta + \frac{\eta}{4}\zeta^2+\frac{1}{H}\zeta\dot{\zeta}+
\frac{1}{4a^2H^2}[-(\vec{\nabla}\zeta)^2 +\nabla^{-2}(\nabla^i\nabla^j(\nabla_i\zeta\nabla_j\zeta))] \nonumber \\
&\;&  + \frac{1}{2a^2H}[\vec{\nabla}\zeta\cdot \vec{\nabla}\chi -\nabla^{-2}(\nabla^i\nabla^j(\nabla_i\zeta\nabla_j\chi))]\,.
\label{redef2}
\eea
Most of these terms involve time and/or spatial derivatives, and hence give negligible contribution deep in the ekpyrotic scaling phase,
when the modes of interest are well outside the Hubble radius. The one possible exception is the $\eta\zeta^2$ term, but this contribution is also suppressed deep in the ekpyrotic
scaling phase, since $\epsilon \approx c^2/2$ is approximately constant and hence $\eta\rightarrow 0$. We can therefore safely ignore the contributions from the field redefinition.

\end{itemize}

\subsection {Summary}

The non-vanishing contributions to the three-point amplitude, given in~(\ref{eps3}),~(\ref{Adoteta}),~(\ref{Azetadotzeta}),~(\ref{Avecnabla}), and~(\ref{Amessyterm}), are
\bea
\nonumber
{\cal A}_{\epsilon^3} &=& \frac{K^2}{32H_0^2}\left(\sum_ik_i^3 - \sum_{i\neq j}k_ik_j^2+2k_1k_2k_3\right)\\
\nonumber
{\cal A}_{\dot{\eta}} &=& - \frac{\pi}{8}\frac{K}{|H_0|} \left( \frac{K}{2}\sum_ik_i^2 - \sum_{i\neq j}k_ik_j^2 +k_1k_2k_3\right)\\
\nonumber
{\cal A}_{\zeta \dot\zeta^2} &=& -\frac{1}{2c^2H_0^2}\sum_{i\neq j}k_i^2k_j^3\,\log(K|t_{\rm end-tran}|)\\
\nonumber
{\cal A}_{\zeta (\vec{\nabla}\zeta)^2} &=& -\frac{1}{12c^2H_0^2} \sum_i k_i^2\sum_jk_j^3 \log(K|t_{\rm end-tran}|) \\
{\cal A}_{\dot{\zeta}\vec{\nabla} \zeta \vec{\nabla} \chi} &=& \frac{K^2}{2c^2H_0^2}\left(\sum_ik_i^3-\frac{3}{2}\sum_{i\neq j}k_ik_j^2+\frac{2}{K^2}\sum_{i\neq j}k_i^2k_j^3 + 4k_1k_2k_3\right)\log(K|t_{\rm end-tran}|)\,.
\label{allAs}
\eea

\end{document}